\documentclass{jfm}
\usepackage{graphicx}
\usepackage{epstopdf, epsfig}

\usepackage{subcaption}

\usepackage{amsmath}
\usepackage{amssymb}
\renewcommand{\Re}{\Rey}
\newcommand{\Rep}{\Rey_{p}} 
\renewcommand{\vec}[1]{\boldsymbol{#1}} 

\shorttitle{Inertial lift in curved ducts}
\shortauthor{B. Harding, Y. M. Stokes and A. L. Bertozzi}
\title{Effect of inertial lift on a spherical particle suspended in flow through a curved duct}

\author{
  Brendan Harding\aff{1}
  \corresp{\email{brendan.harding@adelaide.edu.au}},
  Yvonne M. Stokes\aff{1},
  \and Andrea L. Bertozzi\aff{2}
}

\affiliation{
  \aff{1}School of Mathematical Sciences, The University of Adelaide,
  Adelaide, South Australia 5005, Australia
  \aff{2} Departments of Mathematics and Mechanical and Aerospace Engineering, University of California, 
  Los Angeles, California 90095, USA
}

\begin{document}

\maketitle

\abstract{
We develop a model of the forces on a spherical particle suspended in flow through a curved duct under the assumption that the particle Reynolds number is small.
This extends an asymptotic model of inertial lift force previously developed to study inertial migration in straight ducts.
Of particular interest is the existence and location of stable equilibria within the cross-sectional plane towards which particles migrates.
The Navier--Stokes equations determine the hydrodynamic forces acting on a particle.
A leading order model of the forces within the cross-sectional plane is obtained through the use of a rotating coordinate system and a perturbation expansion in the particle Reynolds number of the disturbance flow.
We predict the behaviour of neutrally buoyant particles at low flow rates and
 examine the variation in focusing position with respect to particle size and bend radius, independent of the flow rate.
In this regime, the lateral focusing position of particles approximately collapses with respect to a dimensionless parameter dependent on three length scales, specifically the particle radius, duct height, and duct bend radius.
Additionally, a trapezoidal shaped cross-section is considered in order to demonstrate how changes in the cross-section design influence the dynamics of particles.
}

\section{Introduction}

Inertial lift force influences the motion of particles suspended in fluid flow through a duct. 
The effect was first demonstrated in the classical experiment of \citet{SegreSilberberg1961} where particles suspended in flow through a (straight) cylindrical pipe were observed to migrate towards an annulus approximately $0.6$ times the radius of the pipe.
This sparked many studies of the simplified scenario of a spherical particle suspended in Poiseuille flow between two plane parallel walls (e.g. \citet{HoLeal1974}, \citet{SchonbergHinch1989} and \citet{Asmolov1999}).
%
Subsequently, particle migration in a circular pipe was investigated at much larger Reynolds numbers ($\Rey$) with studies showing that the radius of the focusing annulus grows with increasing $\Rey$~\citep{MatasMG2004,MatasMG2009}.
In recent years inertial lift has been used in the field of microfluidics for the separation of cells by size and has been extensively studied experimentally \citep{DiCarlo2009,MartelToner2014,Warkianietal2014,GeislingerFranke2014}. 
Much of the behaviour, especially in complex geometries, is still only understood at an empirical level.

In the case of fully enclosed non-circular ducts the inertial lift force is generally much more difficult to approach from an analytical perspective.
Several experimental studies of straight ducts with square or rectangular cross-sections have shown that particles typically tend to migrate towards one of four (stable) equilibrium positions located a finite distance from the centre of each wall \citep{AbbasEtal2014,AminiLeeDiCarlo2014,DiCarloEtal2009,CiftlikEG2013}.
Recent work by \citet{HoodLeeRoper2015} extended the analysis of \citet{SchonbergHinch1989} from the case of Poiseuille flow between two walls to Poiseuille flow through square ducts.
They develop a model of the inertial lift force at small particle Reynolds number which is calculated via the Lorentz reciprocal theorem and utilises the finite element method (FEM) to approximate terms in the expansion that are analytically intractable.
Their results agree well with experimental data and demonstrate that much of the behaviour can be characterised by effects that occur for small particle Reynolds number. 
Their approach was also applied to rectangular ducts where it is shown that the relative size of the attraction zone for equilibria near the side walls becomes increasingly small as the aspect ratio increases, and even disappears entirely for large enough particles~\citep{HoodThesis}. 
Other studies have considered the behaviour at higher Reynolds numbers, in particular \citet{NakagawaEtal2015} used an immersed boundary method to approximate the full Navier--Stokes equations and demonstrate that particles may additionally focus near the corners at high flow rates in agreement with experiments of \citet{MiuraEtal2014}.
These studies also clarify that particles are less focused at high flow rates suggesting that lower flow rates are more useful for cell separation and sorting. 

This paper extends the work of \citet{HoodLeeRoper2015} to the case of curved ducts.
The focusing behaviour of curved and spiral ducts has been largely confined to experimental studies which have shown it to be an effective mechanism to separate particles/cells by size \citep{MartelToner2012,MartelToner2013,RamachandraiahEtal2014,RussomEtal2009,Nivedita2017}.
There have been some studies utilising direct numerical simulation, see for example~\cite{LiuEtal2016} and \cite{OokawaraEtal2006}, in which the inertial lift effect is often added as an external force on the particle via an over-simplified model. 
Curved ducts are well-known to develop a secondary flow consisting of two counter-rotating vortices within the duct cross-section commonly referred to as Dean flow in reference to the early studies of \citet{Dean1927} and \citet{DeanHurst1959}.
The experimental literature demonstrates that size based separation is caused by an interplay between the inertial lift force and the drag forces coming from the secondary fluid flow. 
Here we carefully analyse the Navier--Stokes equations to develop a model of the forces, acting on a particle, within the cross-sectional plane.
The model is then applied to the special case of low flow rate to gain some insight into how the inertial lift force and secondary flow drag interact and lead to focusing positions that differ with respect to particle size. 
Motivated by the experiments of \citet{GuanEtal2013} and \citet{Warkianietal2014} we also consider a curved duct having a trapezoidal cross-section to better understand how the shape of a duct influences the separation of different size particles. 

Microfluidic devices used for particle separation often have a spiral design in order to accommodate the required focusing length relative to the bend radius. 
Although the bend radius in such devices is continuously changing, it has been demonstrated that the slowly changing curvature in such devices has negligible impact on the fluid flow in the sense that it is reasonable to treat the bend radius as being locally constant at each point within the spiral \citep{HardingStokes2018}. 
Therefore, ducts with constant bend radius are considered in this study and, by smoothly interpolating results obtained for several different bend radii (using the same cross-section), we are able to approximate the particle dynamics at any location within a spiral duct.

A leading order model of the relevant cross-sectional forces is developed in Section~\ref{sec:analysis}.
There are several key steps in the case of a curved duct.
The first is the introduction of a rotating reference frame in which the fluid and particle motion is approximately steady. 
This allows us to neglect time dependence of the cross-sectional forces on the particle, but also introduces inertia and Coriolis forces into the Navier--Stokes equations and force model.
The second step is the introduction of disturbance flow variables followed by a non-dimensionalisation of the problem and a perturbation expansion with respect to the particle Reynolds number.
This is similar in spirit to the approach of \citep{HoodLeeRoper2015} for a straight duct.
The Lorentz reciprocal theorem is used to re-express first order forces, which include the inertial lift force, as a volume integral depending on the leading order approximation of the disturbance flow. 
Third, we break up the background flow velocity into its axial and secondary components and use this to further expand the force and torque experienced by the particle into distinct parts.
Lastly, a model for the trajectory of a particle is developed based on its terminal velocity resulting from the forces driving its motion within the plane of the duct cross-section.

In Section~\ref{sec:lfr_limit} we apply the general model from the preceding section to the specific case of particle behaviour in the limit of low flow rate.
This is an interesting case to consider for two reasons.
Firstly it allows us to consider a simplified approximation of the background fluid flow and eliminate dependence on flow velocity so that we may focus on the effect of bend radius and particle size on the focusing behaviour.
In particular, by utilising an expansion of the background flow developed in \citep{Harding2018} we demonstrate that it is sufficient to take a leading order approximation of the axial and secondary components of the flow.
It then becomes clear that the interaction between the secondary flow drag and inertial lift force converges towards a limiting behaviour (for a fixed particle size and duct dimensions) that is different from that obtained for straight ducts.
Second, it allows us to validate the model by demonstrating that the focusing behaviour observed in straight ducts is recovered for large bend radii.
This provides new insights into how the focusing behaviour in curved ducts differs from that in straight ducts. 

In Section~\ref{sec:results} we present and discuss the focusing behaviour resulting from our model for ducts having square, rectangular and trapezoidal cross-sections.
Curved square ducts are not be particularly good at focusing or separating particles but illustrate some interesting dynamics which we comment on briefly.
On the other hand, rectangular and trapezoidal ducts are quite effective at focusing particles and the focusing position depends on the size of the particle and bend radius of the duct.
Furthermore, the behaviour approximately collapses to a single curve when plotted with respect to a dimensionless parameter which describes the relative size of the inertial lift and secondary drag forces.
These results provide insights into recently published experimental results.

\section{Governing equations and perturbation expansion}\label{sec:analysis}

Here we derive the equations for our leading order force model.

\subsection{Governing equations in the lab frame of reference}\label{sec:gov_eqns}

\begin{figure}
\centering
\includegraphics{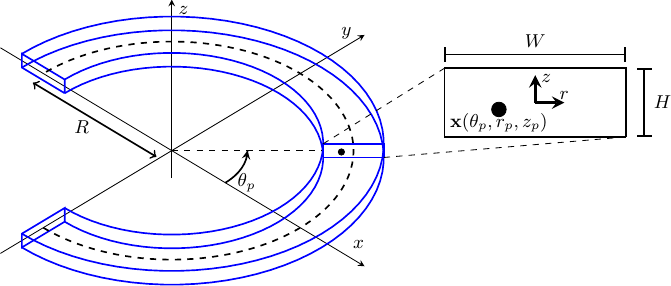}
\caption{
Curved duct with rectangular cross-section containing a spherical particle located at $\mathbf{x}_{p}=\mathbf{x}(\theta_{p},r_{p},z_{p})$. 
The enlarged view of the cross-section around the particle illustrates the origin of the local $r,z$ coordinates at the centre of the duct.
The bend radius $R$ is with respect to the centre-line of the duct and is of modest size for illustration.}
\label{fig:curved_duct}
\end{figure}

Consider a duct which is curved having a constant bend radius and uniform cross-sectional shape, an example of which is depicted in Figure~\ref{fig:curved_duct}.
Let $\vec{x}=(x,y,z)$ denote a Cartesian coordinate system in the \textit{lab} reference frame with the duct positioned so that the $z$ axis is normal to the plane of the bend.
Let the bend radius $R$ be the distance from the origin to the centre of the duct cross-section at any axial/angular position.
We introduce $r,\theta$ such that $r,z$ are coordinates that describe the location within the duct cross-section relative to its centre and $\theta$ is the angle around the bend measured from the $x$-axis, see Figure~\ref{fig:curved_duct}. Note that we do not consider the flow near the inlet/outlet.
With this description $\vec{x}$ is parametrised as
$$
\vec{x}(r,\theta,z)=(R+r)\cos(\theta)\vec{e}_{x}+(R+r)\sin(\theta)\vec{e}_{y}+z\vec{e}_{z} .
$$ 
For example, the point on the centre-line $\vec{x}=R\vec{e}_{x}$ is equivalently described by $(r,\theta,z)=(0,0,0)$.
For a duct having a rectangular cross-section with width $W$ and height $H$ the walls of the duct are taken to be located at $r=\pm W/2$ and $z=\pm H/2$.
For the non-rectangular ducts considered later we take $H$ to be the average height and the origin of the cross-section to be such that for $r=0$ the top and bottom walls are located at $z=\pm H/2$ respectively.
Let $\ell:=H$ be a characteristic minimum width of the duct cross-section noting that the ducts considered in this paper all have $H\leq W$.

Consider a steady fluid flow, through the duct, satisfying the incompressible Navier--Stokes equations.
Letting $\bar{\vec{u}}$ denote the fluid velocity and $\bar{p}$ denote the pressure, we have
\begin{subequations}\label{eqn:ns_bg}
\begin{align}
-\nabla\bar{p}+\mu\nabla^{2}\bar{\vec{u}} &= \rho\left(\bar{\vec{u}}\cdot\nabla\bar{\vec{u}}-\vec{g}\right)  && \text{on $\vec{x}\in\mathcal{D}$} ,\label{eqn:ns_bg_a} \\ 
\nabla\cdot\bar{\vec{u}} &= 0  && \text{on $\vec{x}\in\mathcal{D}$} , \\
\bar{\vec{u}} &= \vec{0} && \text{on $\vec{x}\in\partial\mathcal{D}$} ,
\end{align}
\end{subequations}
where $\rho$ and $\mu$ denote the fluid density and viscosity respectively (both assumed to be constant), $\vec{g}$ is the (constant) acceleration due to gravity, $\mathcal{D}$ denotes the interior of the duct and $\partial\mathcal{D}$ denotes its boundary/surface (i.e. the walls of the duct).
The fluid motion is driven by a steady pressure gradient such that $\bar{p}$ can be decomposed into
$$
\bar{p}(\vec{x}(r,\theta,z)) = -GR\theta+\rho\vec{x}\cdot\vec{g}+\tilde{p}(r,z) ,
$$
where $G$ denotes the drop in pressure per unit arc-length along the centreline of the duct.
Notice that we have assumed gravity does not influence the flow velocity by introducing the hydrostatic pressure $\rho\vec{x}\cdot\vec{g}$ to counteract $\rho\vec{g}$ in \eqref{eqn:ns_bg_a}.
For simplicity our study is restricted to the specific case where gravity acts in the $z$ direction, that is $\vec{g}=-g\vec{e}_{z}$.

The resulting flow, often referred to as Dean flow, has been well-studied for ducts having circular or rectangular cross-sections \citep{DeanHurst1959,Winters1987,Yanase1989,Yamamoto2004}. 
The pressure gradient drives an axial flow from which the inertia of the fluid around the bend then leads to the development of a secondary flow consisting of two counter rotating vortices within the duct cross-section. 
(At high Dean numbers there can exist multiple solutions, some having several vortices, however such flow conditions are far beyond those of practical interest in the context of microfluidics).
Sufficiently far from the inlet/outlet region, the flow velocity components are independent of the angular coordinate (up to their direction) and can be expressed as \begin{equation}\label{eqn:bar_u_expanded}
\bar{\vec{u}}=\bar{u}_{\theta}(r,z)\hat{\vec{e}}_{\theta}+\bar{u}_{r}(r,z)\hat{\vec{e}}_{r}+\bar{u}_{z}(r,z)\vec{e}_{z},
\end{equation}
where $\hat{\vec{e}}_{\theta}:=-\sin(\theta)\vec{e}_{x}+\cos(\theta)\vec{e}_{y}$ and $\hat{\vec{e}}_{r}:=\cos(\theta)\vec{e}_{x}+\sin(\theta)\vec{e}_{y}$.
The $\bar{u}_{\theta}$ component is the axial flow velocity whereas the $\bar{u}_{r},\bar{u}_{z}$ components describe the secondary flow.
Throughout this paper $\bar{\vec{u}}$ and $\bar{p}$ are referred to as the background velocity and pressure respectively.
In the context of studying the dynamics of a particle suspended in flow through a curved duct $\bar{\vec{u}},\bar{p}$ will be treated as being known/prescribed.
Details of the specific approximation of $\bar{\vec{u}}$ used to obtain our results is deferred to Section \ref{sec:lfr_limit} while the treatment throughout the remainder of this section is quite general. 

Consider a single spherical particle within the fluid flow through the duct. 
There are two ways one could analyse the trajectory of a particle within the flow.
If the secondary component of the background flow is relatively strong (i.e. such that inertial lift force will be small compared to fluid drag from the secondary flow) then one might view the particle as behaving like a passive tracer (with respect to the background flow velocity $\bar{\vec{u}}$) which is then perturbed by additional drag (due to the nearby walls) and inertial lift forces.
Alternatively, if the secondary component of the background flow is moderate or weak (i.e. such that inertial lift force is similar in magnitude to drag from the secondary fluid motion) one could consider the particle to be travelling along a streamline of the axial component of the background flow (i.e. streamlines of $\bar{u}_{\theta}\hat{\vec{e}}_{\theta}$) and that its motion is perturbed by the drag forces from the secondary fluid motion, increased drag from the walls, and inertial lift forces.
We consider applications in which the drag forces from the secondary fluid motion is not significantly larger than the inertial lift force and thus take the latter point of view throughout this paper.

Let $a$ denote the radius of the spherical particle, $\rho_{p}$ denote its density (which we assume to be uniform), and $\vec{x}_{p}(t)=(x_{p}(t),y_{p}(t),z_{p}(t))$ denote the location of its centre at time $t$.
It is convenient to parametrise the particle position in cylindrical coordinates as $\vec{x}_{p}(t) = \vec{x}(r_{p}(t),\theta_{p}(t),z_{p}(t))$, see Figure~\ref{fig:curved_duct}.
The particle follows an axial flow streamline (from which it will be perturbed) and thus $\Theta:=\partial\theta_{p}/\partial t$ is constant, $\partial r_{p}/\partial t=\partial z_{p}/\partial t=0$, and $\partial \vec{\Omega}_{p}/\partial t=\vec{0}$ where $\vec{\Omega}_{p}$ denotes the spin of the particle (i.e. the angular velocity with respect to its centre).
The assumption $\partial r_{p}/\partial t=\partial z_{p}/\partial t=0$ may seem counter-intuitive, i.e. we assume that $r_{p},z_{p}$ is fixed only to calculate non-zero forces in these directions such that they cannot remain fixed.
However, this is consistent with the approaches in previous studies of inertial lift forces in uni-directional flows, see for example \citet{SchonbergHinch1989} and \citet{HoodLeeRoper2015}.
Furthermore, it can be shown that any additional effects from particle motion in the $r,z$ directions is much smaller than the inertial lift force and can therefore be neglected (given the other assumptions used here). 
Without loss of generality we can additionally assume that $\theta_{p}=0$ at time $t=0$ such that $\theta_{p}(t)=U_{p}t/(R+r_{p})$ (or equivalently $\Theta=U_{p}/(R+r_{p})$) where $U_{p}$ here denotes the (linear) velocity of the particle.

The presence of the particle alters the fluid flow and we introduce $\vec{u},p$ to denote the fluid velocity and pressure in the presence of the particle.
The fluid motion is again assumed to be governed by the incompressible Navier--Stokes equations, that is, in general
\begin{subequations}\label{eqn:ns_lab}
\begin{align}
-\nabla p+\mu\nabla^{2}\vec{u} &= \rho\left(\frac{\partial \vec{u}}{\partial t}+\vec{u}\cdot\nabla\vec{u}+g\vec{e}_{z}\right) && \text{on $\vec{x}\in\mathcal{F}$} , \label{eqn:ns_lab_a} \\
\nabla\cdot\vec{u}&=0 \quad \text{on $\vec{x}\in\mathcal{F}$} &\quad\vec{u} = \vec{0} \quad& \text{on $\vec{x}\in\partial\mathcal{D}$} , \\ 
\vec{u} &= \frac{\partial\vec{x}_{p}}{\partial t}+\vec{\Omega}_{p}\times(\vec{x}-\vec{x}_{p}) && \text{on $\vec{x}\in\partial\mathcal{F}\backslash\partial\mathcal{D}$} \label{eqn:ns_lab_d} ,
\end{align}
\end{subequations}
where $\mathcal{F}:=\{\vec{x}\in\mathcal{D}:|\vec{x}-\vec{x}_{p}|\geq a\}$ is the fluid domain given the presence of the spherical particle and $\partial\mathcal{F}\backslash\partial\mathcal{D}=\{\vec{x}:|\vec{x}-\vec{x}_{p}|=a\}$ is the surface of the particle.
Note that the moving particle results in unsteady fluid flow in the \textit{lab} frame of reference.
In particular, $\mathcal{F}$ depends on time because$\vec{x}_{p}$ is non-stationary. 
Thus the $\partial \vec{u}/\partial t$ in \eqref{eqn:ns_lab_a} cannot be discarded (as it was for the background flow in \eqref{eqn:ns_bg_a}).
Note that the boundary condition \eqref{eqn:ns_lab_d} has a constant $\vec{\Omega}_{p}$ that denotes the spin (or angular velocity) of the particle (as observed in the \textit{lab} frame). 
Far from the particle we expect $\vec{u},p$ to converge to the background flow $\bar{\vec{u}},\bar{p}$.

The particle motion is driven solely by the force and torque exerted on the particle by the surrounding fluid in addition to the gravitational body force.
The former are obtained by integrating the fluid stresses over the particle surface:
\begin{subequations}\label{eqn:FT_lab}
\begin{align}
\vec{F}(t)&=-m_{p}g\vec{e}_{z}+\int_{|\vec{x}-\vec{x}_{p}|=a} (-\vec{n})\cdot(-p\mathbb{I}+\mu(\nabla\vec{u}+\nabla\vec{u}^{\intercal})) \,dS , \label{eqn:FT_lab_a} \\
\vec{T}(t)&=\int_{|\vec{x}-\vec{x}_{p}|=a} (\vec{x}-\vec{x}_{p})\times\left((-\vec{n})\cdot(-p\mathbb{I}+ \mu(\nabla\vec{u}+\nabla\vec{u}^{\intercal}))\right) \,dS  ,
\end{align}
\end{subequations}
where $m_{p}:=(4/3)\pi a^{3}\rho_{p}$ is the mass of the particle.
Note that we take the normal $\vec{n}$ with respect to the fluid domain $\mathcal{F}$ and thus on the surface of the spherical particle $-\vec{n}$ is the normal vector pointing outwards from the centre of the particle.

\subsection{Equations of motion in a rotating frame of reference}

Although the fluid motion is not steady in the \textit{lab} frame,we view it as steady relative to the motion of the particle centre (assumed to be moving with constant angular velocity $\partial \theta_{p}/\partial t=\Theta$ and fixed $r_{p},z_{p}$).
Therefore, in the particle reference frame (noting that the particle may still have a constant spin) we can eliminate the $\partial \vec{u}/\partial t$ term.

Consider a reference frame  \textit{rotating} about the $z$ axis at the (constant) rate $\partial \theta_{p}/\partial t=:\Theta$. 
We introduce the coordinate system $\vec{x}'=(x',y',z')$ which we parametrise by $(r',\theta',z')$ such that
$$
\vec{x}'(r',\theta',z') = (R+r')\cos(\theta')\vec{e}_{x'}+(R+r')\sin(\theta')\vec{e}_{y'}+z'\vec{e}_{z'} .
$$
The unit vectors $\vec{e}_{x'},\vec{e}_{y'},\vec{e}_{z'}$ in the  \textit{rotating} frame are related to $\vec{e}_{x},\vec{e}_{y},\vec{e}_{z}$ in the \textit{lab} reference frame by
$$
\vec{e}_{x'}=\cos(\theta_{p})\vec{e}_{x}+\sin(\theta_{p})\vec{e}_{y} ,\quad
\vec{e}_{y'}=-\sin(\theta_{p})\vec{e}_{x}+\cos(\theta_{p})\vec{e}_{y} ,\quad
\vec{e}_{z'}=\vec{e}_{z} .
$$
or conversely
$$
\vec{e}_{x}=\cos(\theta_{p})\vec{e}_{x'}-\sin(\theta_{p})\vec{e}_{y'} ,\quad
\vec{e}_{y}=\sin(\theta_{p})\vec{e}_{x'}+\cos(\theta_{p})\vec{e}_{y'} ,\quad
\vec{e}_{z}=\vec{e}_{z'} .
$$
The angular velocity of the \textit{rotating} reference frame (relative to the \textit{lab} frame) is 
\begin{equation}\label{eqn:Theta_def}
\Theta\vec{e}_{z} := \frac{\partial\theta_{p}}{\partial t}\vec{e}_{z} = \frac{U_{p}}{R+r_{p}}\vec{e}_{z} .
\end{equation}

Let $\vec{x}_{p}^{\prime}=(x_{p}^{\prime},y_{p}^{\prime},z_{p}^{\prime})$ be the centre of the particle in the \textit{rotating} frame of reference. 
It is always the case that $x_{p}^{\prime}=R+r_{p}$, $y_{p}^{\prime}=0$ and $z_{p}^{\prime}=z_{p}$.
Consequently, the particle velocity in the \textit{rotating} reference frame is
$$
\frac{\partial \vec{x}_{p}^{\prime}}{\partial t} 
= \frac{\partial x_{p}^{\prime}}{\partial t}\vec{e}_{x'} + \frac{\partial z_{p}^{\prime}}{\partial t}\vec{e}_{z'}
= \frac{\partial r_{p}}{\partial t}\vec{e}_{x'} + \frac{\partial z_{p}}{\partial t}\vec{e}_{z'} 
= 0 .
$$
The particle spin as viewed from the \textit{rotating} reference frame is reduced by the solid body rotation of the coordinate system and is therefore denoted
$$
\vec{\Omega}_{p}^{\prime}:=\vec{\Omega}_{p}-\Theta\vec{e}_{z'} .
$$

Let $\vec{u}'(\vec{x}',t)$ denote the velocity of a fluid parcel in the \textit{rotating} frame which is related to the velocity $\vec{u}(\vec{x},t)$ in the \textit{lab} frame via
\begin{equation*}
\vec{u}(\vec{x},t) =  \vec{u}'(\vec{x}',t) + \Theta(\vec{e}_{z'}\times\vec{x}') .
\end{equation*}
Furthermore, the acceleration of a fluid parcel in the two frames are related via
\begin{equation*}
\frac{d \vec{u}}{d t} =  \frac{D \vec{u}'}{D t} + 2\Theta(\vec{e}_{z'}\times\vec{u}') + \Theta^{2}(\vec{e}_{z'}\times(\vec{e}_{z'}\times\vec{x}')) ,
\end{equation*}
where $D\vec{u}'/Dt:=\partial \vec{u}'/\partial t + \vec{u}'\cdot\nabla'\vec{u}'$ denotes the material derivative.
Letting $p'$ denote the fluid pressure in the rotating frame 
(which is such that $p'(\vec{x}',t)=p(\vec{x},t)$ and thus $\nabla' p'=\nabla p$), 
it follows that the equations of motion for the fluid in the \textit{rotating} reference frame are
\begin{subequations}\label{eqn:ns_rotating}
\begin{align}
-\nabla'p'+\mu\nabla^{\prime\,2}\vec{u}' &= \rho\bigg(\frac{\partial \vec{u}'}{\partial t}+\vec{u}'\cdot\nabla'\vec{u}'+2\Theta(\vec{e}_{z'}\times\vec{u}') &&\notag \\
&\qquad\qquad+\Theta^{2}(\vec{e}_{z'}\times(\vec{e}_{z'}\times\vec{x}'))+g\vec{e}_{z} \bigg) && \text{on $\vec{x}'\in\mathcal{F}'$} , \label{eqn:ns_rotating_a} \\
\nabla'\cdot\vec{u}' &= 0 && \text{on $\vec{x}'\in\mathcal{F}'$} , \\
\vec{u}' &= -\Theta(\vec{e}_{z'}\times\vec{x}') && \text{on $\vec{x}'\in\partial\mathcal{D}$} , \label{rotatingwall}\\
\vec{u}' &= \vec{\Omega}_{p}^{\prime}\times(\vec{x}'-\vec{x}_{p}^{\prime}) && \text{on $\vec{x}'\in\partial\mathcal{F}'\backslash\partial\mathcal{D}$} , \label{eqn:ns_rotating_d}
\end{align}
\end{subequations}
where $\mathcal{F}':=\{\vec{x}'\in\mathcal{D}:|\vec{x}'-\vec{x}_{p}^{\prime}|\geq a\}$ denotes the fluid domain with respect to the \textit{rotating} reference frame. 
The boundary condition (\ref{rotatingwall}) arises from the no slip condition on the walls, rotating with angular velocity $-\Theta\vec{e}_{z}$.
Notice that, expressed in this frame, the fluid flow can be considered to be steady since $\mathcal{F}'$ is stationary and each of $\Theta,\vec{\Omega}_{p}^{\prime},g$ are constant.
Consequently $\partial \vec{u}'/\partial t$ is zero and can be dropped from \eqref{eqn:ns_rotating_a}.
Observe that it is important for gravity to acts in the $\vec{e}_{z}$ direction since otherwise the relative direction of gravity in the rotating frame would be dependent on $\theta_{p}(t)$.

The force and torque on the particle need to be expressed in terms of the \textit{rotating} reference frame as well.
Let $\vec{F}',\vec{T}'$ denote the force and torque on the particle in the \textit{rotating} reference frame, respectively. 
Then one has
\begin{subequations}\label{eqn:FT_relation}
\begin{align}
\vec{F} &= \vec{F}' +m_{p}\Theta^{2}(\vec{e}_{z'}\times(\vec{e}_{z'}\times\vec{x}_{p}^{\prime})) , \\
\vec{T} &= \vec{T}' +I_{p}\Theta(\vec{e}_{z'}\times\vec{\Omega}_{p}^{\prime})  , 
\end{align}
\end{subequations}
where $m_{p}:=(4/3)\pi a^{3}\rho_{p}$ and $I_{p}:=(2/5)a^{2}m_{p}$ are the mass and moment of inertia of the particle respectively. 
Note there is no Coriolis like force since the particle velocity is taken to be zero in the \textit{rotating} reference frame, nor is there an Euler force since $\Theta$ is assumed to be constant.
Thus, upon re-expressing \eqref{eqn:FT_lab} in terms of $p',\vec{u}'$, and then combining with \eqref{eqn:FT_relation} one obtains
\begin{subequations}\label{eqn:FT_rotating}
\begin{align}
\vec{F}' &= -m_{p}g\vec{e}_{z'}-m_{p}\Theta^{2}(\vec{e}_{z'}\times(\vec{e}_{z'}\times\vec{x}_{p}^{\prime})) \notag \\
&\quad +\int_{|\vec{x}'-\vec{x}_{p}^{\prime}|=a} (-\vec{n})\cdot(-p'\mathbb{I}+ \mu(\nabla'\vec{u}'+{\nabla'\vec{u}^{\prime}}^{\intercal})) \,dS' , \\
\vec{T}' &= -I_{p}\Theta(\vec{e}_{z'}\times\vec{\Omega}_{p}^{\prime}) \notag \\
&\quad +\int_{|\vec{x}'-\vec{x}_{p}^{\prime}|=a} (\vec{x}'-\vec{x}_{p}^{\prime})\times\left((-\vec{n})\cdot(-p'\mathbb{I}+ \mu(\nabla'\vec{u}'+{\nabla'\vec{u}^{\prime}}^{\intercal}))\right) \,dS' .
\end{align}
\end{subequations}
Note that $-m_{p}\Theta^{2}(\vec{e}_{z'}\times(\vec{e}_{z'}\times\vec{x}_{p}^{\prime}))$ provides the centripetal force experienced by the particle due to its rotational motion as it travels through the curved duct.
 
We also need the background flow in the \textit{rotating} reference frame.
Since the background flow is steady in time and independent of angular position then, letting $\bar{p}',\bar{\vec{u}}'$ denote pressure and velocity of the background fluid flow respectively, they are related to those in the stationary/lab frame via
\begin{equation}\label{eqn:bg_rotating}
\bar{\vec{u}}'(\vec{x}')=\bar{\vec{u}}(\vec{x}')-\Theta\vec{e}_{z'}\times\vec{x}' , \qquad \bar{p}'(\vec{x}')=\bar{p}(\vec{x}) .
\end{equation}
We are now equipped to consider the disturbance flow in the \textit{rotating} reference frame, that is the difference between flow $\vec{u}',p'$ in the presence of the particle and the flow $\bar{\vec{u}}',\bar{p}'$ in the absence of the particle.

\subsection{Disturbance flow in the rotating frame}

In the \textit{rotating} reference frame we introduce the disturbance velocity and pressure
$$
\vec{v}'=\vec{u}'-\bar{\vec{u}}' , \qquad q'=p'-\bar{p}' ,
$$
respectively.
Substituting $\vec{u}'=\vec{v}'+\bar{\vec{u}}'=\vec{v}'+\bar{\vec{u}}-\Theta(\vec{e}_{z'}\times\vec{x}')$ and $p'=q'+\bar{p}'$ into \eqref{eqn:ns_rotating} 
one obtains
\begin{subequations}\label{eqn:ns_disturbance}
\begin{align}
-\nabla'q'+\mu\nabla^{\prime\,2}\vec{v}' &= \rho\big((\vec{v}'+\bar{\vec{u}}-\Theta(\vec{e}_{z'}\times\vec{x}'))\cdot\nabla'\vec{v}'&& \notag\\
&\qquad\qquad+\vec{v}'\cdot\nabla'\bar{\vec{u}}+\Theta(\vec{e}_{z'}\times\vec{v}')\big) && \text{on $\vec{x}'\in\mathcal{F}'$} , \\
\nabla'\cdot\vec{v}' &= 0 && \text{on $\vec{x}'\in\mathcal{F}'$} , \\
\vec{v}' &= \vec{0} && \text{on $\vec{x}'\in\partial\mathcal{D}$} , \\
\vec{v}' &= -\bar{\vec{u}}+\Theta(\vec{e}_{z'}\times\vec{x}')+\vec{\Omega}_{p}^{\prime}\times(\vec{x}'-\vec{x}_{p}^{\prime}) && \text{on $\vec{x}'\in\partial\mathcal{F}'\backslash\partial\mathcal{D}$} , \label{eqn:ns_disturbance_d}
\end{align}
\end{subequations}
where we have additionally used the fact that
\begin{itemize}
\item[(i)] the background flow $\bar{\vec{u}},\bar{p}$ satisfies \eqref{eqn:ns_bg},
\item[(ii)] for any $\vec{w}$ one has $\vec{w}\cdot\nabla'(\vec{e}_{z'}\times\vec{x}')=\vec{e}_{z'}\times\vec{w}$, 
\item[(iii)] 
using \eqref{eqn:bar_u_expanded} it can be shown $(\vec{e}_{z'}\times\vec{x}')\cdot\nabla'\bar{\vec{u}}=\vec{e}_{z'}\times\bar{\vec{u}}$,
\item[(iv)] for $\vec{x}'\in\partial\mathcal{D}$ one has $\bar{\vec{u}}=\vec{0}$.
\end{itemize}
Similarly the force and torque can be expressed in terms of the disturbance flow as
\begin{subequations}\label{eqn:FT_disturbance}
\begin{align}
\vec{F}' &= -m_{p}g\vec{e}_{z}-m_{p}\Theta^{2}(\vec{e}_{z'}\times(\vec{e}_{z'}\times\vec{x}_{p}^{\prime})) + 
\int_{|\vec{x}'-\vec{x}_{p}^{\prime}|=a} (-\vec{n})\cdot(-\bar{p}\mathbb{I}+\mu( \nabla'\bar{\vec{u}}+{\nabla'\bar{\vec{u}}}^{\intercal})) \,dS'  \notag \\
&\quad+\int_{|\vec{x}'-\vec{x}_{p}^{\prime}|=a} (-\vec{n})\cdot(-q'\mathbb{I}+ \mu(\nabla'\vec{v}'+{\nabla'\vec{v}'}^{\intercal})) \,dS' , \label{eqn:F_disturbance} \\
\vec{T}' &= -I_{p}\Theta(\vec{e}_{z'}\times\vec{\Omega}_{p}^{\prime}) +
\int_{|\vec{x}'-\vec{x}_{p}^{\prime}|=a} (\vec{x}'-\vec{x}_{p}^{\prime})\times\left((-\vec{n})\cdot(-\bar{p}\mathbb{I}+\mu( \nabla'\bar{\vec{u}}+{\nabla\bar{\vec{u}}}^{\intercal}))\right) \,dS' \notag \\
&\quad+\int_{|\vec{x}'-\vec{x}_{p}^{\prime}|=a} (\vec{x}'-\vec{x}_{p}^{\prime})\times\left((-\vec{n})\cdot(-q'\mathbb{I}+ \mu(\nabla'\vec{v}'+{\nabla'\vec{v}'}^{\intercal}))\right) \,dS' , \label{eqn:T_disturbance}
\end{align}
\end{subequations}
respectively (noting that $\nabla'(\vec{e}_{z'}\times\vec{x}')+{\nabla'(\vec{e}_{z'}\times\vec{x}')}^{\intercal}=0$).
Because $\bar{\vec{u}}',\bar{p}'$ are well-defined on the interior of the particle then one may use the divergence theorem and \eqref{eqn:ns_bg_a} to obtain
\begin{multline}
\int_{|\vec{x}'-\vec{x}_{p}^{\prime}|= a} (-\vec{n})\cdot(-\bar{p}\mathbb{I}+\mu( \nabla'\bar{\vec{u}}+{\nabla'\bar{\vec{u}}}^{\intercal})) \,dS' \\
= \rho\int_{|\vec{x}'-\vec{x}_{p}^{\prime}|\leq a} \bar{\vec{u}}\cdot\nabla'\bar{\vec{u}}+g\vec{e}_{z} \,dV' 
= \frac{4}{3}\pi a^{3}\rho g\vec{e}_{z}+\rho\int_{|\vec{x}'-\vec{x}_{p}^{\prime}|\leq a} \bar{\vec{u}}\cdot\nabla'\bar{\vec{u}} \,dV' , \label{eqn:F_bgc}
\end{multline}
recalling that $\vec{n}$ points in to the centre of the particle.
Thus the contribution from the background flow to the force is essentially just the inertia of the fluid that would otherwise take up the volume occupied by the particle. 
For a neutrally buoyant particle (i.e. $\rho_{p}=\rho$) travelling with a velocity close to that of the axial velocity of the background flow we would then expect $m_{p}\Theta^{2}(\vec{e}_{z'}\times(\vec{e}_{z'}\times\vec{x}_{p}^{\prime}))$ to be close to the inertia of the displaced fluid so that the difference between the two is negligible.
If the particle has lower or higher density than the fluid then the net effect of the two terms is a force directed towards the inside or outside wall respectively.
Similarly, for the background contribution to the torque, it may be shown that 
\begin{multline}\label{eqn:T_bgc}
\int_{|\vec{x}'-\vec{x}_{p}^{\prime}|=a} (\vec{x}'-\vec{x}_{p}^{\prime})\times\left((-\vec{n})\cdot(-\bar{p}\mathbb{I}+ \mu(\nabla'\bar{\vec{u}}+{\nabla'\bar{\vec{u}}}^{\intercal}))\right) \,dS' \\
= \rho\int_{|\vec{x}'-\vec{x}_{p}^{\prime}|\leq a} (\vec{x}'-\vec{x}_{p}^{\prime})\times(\bar{\vec{u}}\cdot\nabla'\bar{\vec{u}}) \,dV'  .
\end{multline}
At first glance this may appear like a misuse of the divergence theorem, however it indeed holds by first re-arranging the order of the cross and dot products so that the integrand has the form $-\vec{n}\cdot(\ast)$ after which the divergence theorem can be applied and then further re-arrangement and use of \eqref{eqn:ns_bg_a} leads to \eqref{eqn:T_bgc}. 
In any case this term is sufficiently small that it can be neglected for the purpose of estimating the inertial lift force to leading order as will be demonstrated upon non-dimensionalising our equations.

\subsection{Non-dimensionalisation of the disturbance flow} 

Let us now non-dimensionalise the variables and equations of motion (working in the \textit{rotating} reference frame).
Spatial variables are non-dimensionalised using the radius of the particle $a$ and velocity variables are non-dimensionalised via the characteristic velocity $U:=U_{m}a/\ell$ where $U_{m}$ denotes a characteristic velocity of the background flow $\bar{\vec{u}}$, which we take to be the value of its axial component at $r=\theta=z=0$.
This is essentially the same scale as in \citet{HoodLeeRoper2015}. 
The non-dimensionalisation of the remaining variables follow from these via the standard approach for flows in which viscous stresses are dominant.
Thus we introduce dimensionless variables denoted by $\hat\cdot$, 
\begin{align*}
\vec{x}'&=a\hat{\vec{x}}' , & \vec{v}'&=(U_{m}a/\ell)\hat{\vec{v}}' , & q' &= (\mu U_{m}/\ell)\hat{q}' , \\
\vec{x}_{p}^{\prime}&=a\hat{\vec{x}}_{p}^{\prime} , & \bar{\vec{u}}&=(U_{m}a/\ell)\hat{\bar{\vec{u}}} , & \bar{p} &= (\mu U_{m}/\ell)\hat{\bar{p}} , \\
t &= (\ell/U_{m})\hat{t} , & \vec{\Omega}_{p}^{\prime} &= (U_{m}/\ell)\hat{\vec{\Omega}}_{p}^{\prime} , & \Theta &= (U_{m}/\ell)\hat{\Theta} , \\
\nabla' &= a^{-1}\hat{\nabla}' , & g &= (U_{m}^{2}a/\ell^{2})\hat{g} . && 
\end{align*}
With these scalings, the dimensionless form of \eqref{eqn:ns_disturbance} is
\begin{subequations}\label{eqn:ns_dimless}
\begin{align}
-\hat{\nabla}'\hat{q}'+\hat{\nabla}^{\prime\,2}\hat{\vec{v}}' &= \Rep\big((\hat{\vec{v}}'+\hat{\bar{\vec{u}}}-\hat{\Theta}(\vec{e}_{z'}\times\hat{\vec{x}}'))\cdot\hat{\nabla}'\hat{\vec{v}}'&& \notag\\
&\qquad\qquad\quad+\hat{\vec{v}}'\cdot\hat{\nabla}'\hat{\bar{\vec{u}}}+\hat{\Theta}(\vec{e}_{z'}\times\hat{\vec{v}}')\big) && \text{on $\hat{\vec{x}}'\in\hat{\mathcal{F}}'$} , \label{eqn:ns_dimless_a} \\
\hat{\nabla}'\cdot\hat{\vec{v}}' &= 0 && \text{on $\hat{\vec{x}}'\in\hat{\mathcal{F}}'$} , \\
\hat{\vec{v}}' &= \vec{0} && \text{on $\hat{\vec{x}}'\in\partial\hat{\mathcal{D}}$} , \\
\hat{\vec{v}}' &= -\hat{\bar{\vec{u}}}+\hat{\Theta}(\vec{e}_{z'}\times\hat{\vec{x}}')+\hat{\vec{\Omega}}_{p}^{\prime}\times(\hat{\vec{x}}'-\hat{\vec{x}}_{p}^{\prime}) && \text{on $\hat{\vec{x}}'\in\partial\hat{\mathcal{F}}'\backslash\partial\hat{\mathcal{D}}$} , \label{eqn:ns_dimless_d}
\end{align}
\end{subequations}
where $\hat{\mathcal{F}}':=\{\hat{\vec{x}}':a\hat{\vec{x}}'\in\mathcal{F}'\}$ and $\hat{\mathcal{D}}':=\{\hat{\vec{x}}':a\hat{\vec{x}}'\in\mathcal{D}'\}$ denote the rescaled domains and 
\begin{equation}\label{eqn:Re_p}
\Rep := \frac{\rho}{\mu} U a = \frac{\rho}{\mu}U_{m}\frac{a^{2}}{\ell} ,
\end{equation}
is the particle Reynolds number.  
Writing the duct Reynolds number, $\Rey:=(\rho/\mu)U_{m}\ell$, we have $\Rep=\Rey(a/\ell)^{2}$, so that the particle Reynolds number can be small even if the duct Reynolds number is not (note that $a<\ell/2$ is necessary for the particle to fit in the duct and $a\lesssim \ell/10$ is typical).

The forces on the particle in the rotating frame must also be non-dimensionalised.
Noting that the dimensional force $\vec{F}'$ and torque $\vec{T}'$ are
\begin{equation*}
\vec{F}'=m_{p}\frac{\partial^{2} \vec{x}_{p}^{\prime}}{\partial t^{2}} 
\quad\text{and}\quad \vec{T}'=I_{p}\frac{\partial \vec{\Omega}_{p}^{\prime}}{\partial t} ,
\end{equation*}
we introduce the dimensionless quantities $\hat{\vec{F}}',\hat{\vec{T}}'$ defined by
\begin{equation*}
\hat{\vec{F}}':=\frac{\ell^{2}}{\rho U_{m}^{2}a^{4}}\vec{F}'=\frac{\rho_{p}}{\rho}\frac{4\pi}{3}\frac{\partial^{2} \hat{\vec{x}}_{p}^{\prime}}{\partial \hat{t}^{2}}
,\quad\text{and}\quad 
\hat{\vec{T}}':=\frac{\ell^{2}}{\rho U_{m}^{2}a^{5}}\vec{T}' =\frac{\rho_{p}}{\rho}\frac{8\pi}{15}\frac{\partial \hat{\vec{\Omega}}_{p}^{\prime}}{\partial \hat{t}} .
\end{equation*}
Noting that $\rho U_{m}^{2}a^{4}/\ell^{2} = \Rep\mu U_{m}a^{2}/\ell$, one has from \eqref{eqn:F_disturbance}
\begin{multline}\label{eqn:force_dimless}
\hat{\vec{F}}'=-\frac{\rho_{p}-\rho}{\rho}\frac{4\pi}{3}\hat{g}\vec{e}_{z}
-\frac{\rho_{p}}{\rho}\frac{4\pi}{3}\hat{\Theta}^{2}(\vec{e}_{z'}\times(\vec{e}_{z'}\times\hat{\vec{x}}_{p}^{\prime}))
+\int_{|\hat{\vec{x}}'-\hat{\vec{x}}_{p}^{\prime}|<1} \hat{\bar{\vec{u}}}\cdot\hat{\nabla}'\hat{\bar{\vec{u}}} \,d\hat{V}'   \\
+\frac{1}{\Rep}\int_{|\hat{\vec{x}}'-\hat{\vec{x}}_{p}^{\prime}|=1} (-\vec{n})\cdot(-\hat{q}'\mathbb{I}+ \hat{\nabla}'\hat{\vec{v}}'+{\hat{\nabla}'\hat{\vec{v}}'}^{\intercal}) \,d\hat{S}' ,
\end{multline}
and similarly from \eqref{eqn:T_disturbance}
\begin{multline}\label{eqn:torque_dimless}
\hat{\vec{T}}' = -\frac{\rho_{p}}{\rho}\frac{8\pi}{15}\hat{\Theta}(\vec{e}_{z'}\times\hat{\vec{\Omega}}_{p}^{\prime}) 
+\int_{|\hat{\vec{x}}'-\hat{\vec{x}}_{p}^{\prime}|<1} (\hat{\vec{x}}'-\hat{\vec{x}}_{p}^{\prime})\times\left(\hat{\bar{\vec{u}}}\cdot\hat{\nabla}'\hat{\bar{\vec{u}}}\right) \,d\hat{V}'  \\
+\frac{1}{\Rep}\int_{|\hat{\vec{x}}'-\hat{\vec{x}}_{p}^{\prime}|=1} (\hat{\vec{x}}'-\hat{\vec{x}}_{p}^{\prime})\times\left((-\vec{n})\cdot(-\hat{q}'\mathbb{I}+ \hat{\nabla}'\hat{\vec{v}}'+{\hat{\nabla}'\hat{\vec{v}}'}^{\intercal}\right) \,d\hat{S}' .
\end{multline}

At this stage, given $a,\hat{\vec{x}}_{p}^{\prime},\hat{\Theta},\hat{\vec{\Omega}}_{p}^{\prime}$ and an approximation of $\hat{\bar{\vec{u}}}$ (for a desired duct shape/size and flow rate), one could solve the non-linear problem \eqref{eqn:ns_dimless} and then determine the resulting force and torque on the particle via \eqref{eqn:force_dimless} and \eqref{eqn:torque_dimless} respectively.
In the absence of a secondary component of the background flow the inertial lift force only becomes the dominant force once the particle has reached `terminal' velocity and spin in relation to its main axial motion. 
As such, it is typical to determine $\hat{\Theta},\hat{\vec{\Omega}}_{p}^{\prime}$ such that $\hat{\vec{T}}'$ and the axial component of $\hat{\vec{F}}'$ (i.e. $\hat{\vec{F}}'\cdot\vec{e}_{y'}$) are negligible through an iterative process.
In the simpler case of flow through a straight duct, the remaining components of $\hat{\vec{F}}'$ provide the inertial lift force which leads to drift/migration within the cross-section.
However for curved ducts, the secondary component of the background flow results in $\hat{\vec{F}}'$  being influenced by the drag force from the secondary flow motion. 
A location in the cross-section where the net force from both effects is zero is an equilibrium to/from which a particle may be attracted/repelled depending on the eigenvalues of the Jacobian of the net force (as a function of the particle position within the cross-section, i.e. $\hat{\vec{x}}_{p}^{\prime}$).

In the case of a straight duct, the the inertial lift force can be approximated from a perturbation expansion of the disturbance flow for small $\Rep$ \citep{HoodLeeRoper2015}.
Through this expansion, the non-linear equation \eqref{eqn:ns_dimless} can be broken up into a sequence of linear Stokes problems that are more easily solved. 
Furthermore, the resulting decomposition informs the way in which inertial lift influences the particle motion, particularly in our case where the problem is further complicated by the secondary fluid motion.
The remainder of this section describes how the approach of \citet{HoodLeeRoper2015} is adopted to the curved duct geometry, and additionally, how separating the background flow into axial and secondary components allows us to separate and identify the terms that influence particle motion within the cross-section.

\subsection{Perturbation expansion of the disturbance flow and forces on the particle}\label{sec:Rep_expansion}

We consider a perturbation expansion of the disturbance flow in powers of the particle Reynolds number:
\begin{equation}\label{eqn:vq_pert}
\hat{\vec{v}}'=\vec{v}_{0} + \Rep\vec{v}_{1} + O(\Rep^{2}) , 
\qquad
\hat{q}'=q_{0} + \Rep q_{1} + O(\Rep^{2}) .
\end{equation}
Note that carets are dropped from the perturbation variables since these will always be taken to be dimensionless quantities.
Substituting \eqref{eqn:vq_pert} into \eqref{eqn:ns_dimless} and matching powers of $\Rep$, one obtains the zeroth order equations
\begin{subequations}\label{eqn:perturb_eqn0}
\begin{align}
-\hat{\nabla}'q_{0} + \hat{\nabla}^{\prime\,2}\vec{v}_{0} &= \vec{0} , &&\text{on $\hat{\vec{x}}'\in\hat{\mathcal{F}}'$} , \label{eqn:perturb_eqn0_a} \\
\hat{\nabla}'\cdot\vec{v}_{0} &= 0 , &&\text{on $\hat{\vec{x}}'\in\hat{\mathcal{F}}'$} , \label{eqn:perturb_eqn0_b} \\
\vec{v}_{0} &= \vec{0} , \qquad &&\text{on $\hat{\vec{x}}'\in\partial\hat{\mathcal{D}}'$} , \label{eqn:perturb_eqn0_c} \\
\vec{v}_{0} &= \hat{\vec{\Omega}}_{p}^{\prime}\times(\hat{\vec{x}}'-\hat{\vec{x}}_{p}^{\prime}) +  \hat{\Theta}(\vec{e}_{z'}\times\hat{\vec{x}}^{\prime}) -\hat{\bar{\vec{u}}} , \quad &&\text{on $\hat{\vec{x}}'\in\partial\hat{\mathcal{F}}'\backslash\partial\hat{\mathcal{D}}'$} ,  \label{eqn:perturb_eqn0_d}
\end{align}
\end{subequations}
and the first order equations
\begin{subequations}\label{eqn:perturb_eqn1}
\begin{align}
-\hat{\nabla}'q_{1} + \hat{\nabla}^{\prime\,2}\vec{v}_{1} &= 
\hat{\Theta}(\vec{e}_{z'}\times\vec{v}_{0}) + \vec{v}_{0}\cdot\hat{\nabla}'\hat{\bar{\vec{u}}} && \notag \\
&\quad+ (\vec{v}_{0}+\hat{\bar{\vec{u}}}-\hat{\Theta}(\vec{e}_{z'}\times\hat{\vec{x}}'))\cdot\hat{\nabla}'\vec{v}_{0} , &&\text{on $\hat{\vec{x}}'\in\hat{\mathcal{F}}'$} , \label{eqn:perturb_eqn1_a} \\
\hat{\nabla}'\cdot\vec{v}_{1} &= 0 , &&\text{on $\hat{\vec{x}}'\in\hat{\mathcal{F}}'$} , \\
\vec{v}_{1} &= \vec{0} , &&\text{on $\hat{\vec{x}}'\in\partial\hat{\mathcal{D}}'$} , \\
\vec{v}_{1} &= \vec{0} , &&\text{on $\hat{\vec{x}}'\in\partial\hat{\mathcal{F}}'\backslash\partial\hat{\mathcal{D}}'$} . \label{eqn:perturb_eqn1_d}
\end{align}
\end{subequations}

Equations \eqref{eqn:perturb_eqn0} and \eqref{eqn:perturb_eqn1} are similar to those obtained for the straight duct (see \citet{HoodLeeRoper2015}) but with a few key differences.
Firstly, the right side of \eqref{eqn:perturb_eqn1_a} has additional terms introduced in the change to a rotating coordinate system.
Secondly, the background velocity $\hat{\bar{\vec{u}}}$ is no longer a simple Poiseuille flow through the duct.
Lastly, the boundary condition \eqref{eqn:ns_dimless_d} is captured entirely in \eqref{eqn:perturb_eqn0_d} so that $\vec{v}_{1}=\vec{0}$ in \eqref{eqn:perturb_eqn1_d}.

The perturbation expansion \eqref{eqn:vq_pert} can also be substituted into \eqref{eqn:force_dimless} and \eqref{eqn:torque_dimless} to obtain the expansion
\begin{subequations}\label{eqn:FT_pert}
\begin{align}
\hat{\vec{F}}' &= \Rep^{-1}\vec{F}_{-1} + \vec{F}_{0} + O(\Rep) , \\
\hat{\vec{T}}' &= \Rep^{-1}\vec{T}_{-1} + \vec{T}_{0} + O(\Rep) ,
\end{align}
\end{subequations}
where
\begin{subequations}
\begin{align}
\vec{F}_{-1} :=\,&
\int_{|\hat{\vec{x}}'-\hat{\vec{x}}_{p}^{\prime}|=1} (-\vec{n})\cdot\left(-q_{0}\mathbb{I}+ \hat{\nabla}'\vec{v}_{0}+{\hat{\nabla}'\vec{v}_{0}}^{\intercal}\right) \,d\hat{S}'  , \label{eqn:Fm1} \\
\vec{F}_{0} :=\,& -\frac{\rho_{p}-\rho}{\rho}\frac{4\pi}{3}\hat{g}\vec{e}_{z'}
-\frac{\rho_{p}}{\rho}\frac{4\pi}{3} \hat{\Theta}^{2}(\vec{e}_{z'}\times(\vec{e}_{z'}\times\hat{\vec{x}}_{p}^{\prime})) 
+\int_{|\hat{\vec{x}}'-\hat{\vec{x}}_{p}^{\prime}|<1} \hat{\bar{\vec{u}}}\cdot\hat{\nabla}'\hat{\bar{\vec{u}}} \,d\hat{V}' \notag \\
&+\int_{|\hat{\vec{x}}'-\hat{\vec{x}}_{p}^{\prime}|=1} (-\vec{n})\cdot\left(-q_{1}\mathbb{I}+ \hat{\nabla}'\vec{v}_{1}+{\hat{\nabla}'\vec{v}_{1}}^{\intercal}\right) \,d\hat{S}'  , \label{eqn:F0}
\end{align}
\end{subequations}
and similarly
\begin{subequations}
\begin{align}
\vec{T}_{-1} :=\,& 
\int_{|\hat{\vec{x}}'-\hat{\vec{x}}_{p}^{\prime}|=1} (\hat{\vec{x}}'-\hat{\vec{x}}_{p}^{\prime})\times\left((-\vec{n})\cdot\left(-q_{0}\mathbb{I}+ \hat{\nabla}'\vec{v}_{0}+{\hat{\nabla}'\vec{v}_{0}}^{\intercal}\right)\right) \,d\hat{S}' , \label{eqn:Tm1} \\
\vec{T}_{0} :=\,& -\frac{\rho_{p}}{\rho}\frac{8\pi}{15}\hat{\Theta}(\vec{e}_{z'}\times\hat{\vec{\Omega}}_{p}^{\prime}) 
+\int_{|\hat{\vec{x}}'-\hat{\vec{x}}_{p}^{\prime}|<1} (\hat{\vec{x}}'-\hat{\vec{x}}_{p}^{\prime})\times\left(\hat{\bar{\vec{u}}}\cdot\hat{\nabla}'\hat{\bar{\vec{u}}}\right) \,d\hat{V}' \notag \\
&+\int_{|\hat{\vec{x}}'-\hat{\vec{x}}_{p}^{\prime}|=1} (\hat{\vec{x}}'-\hat{\vec{x}}_{p}^{\prime})\times\left((-\vec{n})\cdot\left(-q_{1}\mathbb{I}+ \hat{\nabla}'\vec{v}_{1}+{\hat{\nabla}'\vec{v}_{1}}^{\intercal}\right)\right) \,d\hat{S}' . \label{eqn:T0}
\end{align}
\end{subequations}

In the case of a straight duct \citet{HoodLeeRoper2015} showed that the integral over the stress from $q_{1},\vec{v}_{1}$ can be computed without explicitly solving for these terms by utilising the Lorentz reciprocal theorem.
The same result holds in the case of a curved duct, specifically it is straightforward to show
\begin{multline}\label{eqn:F0_reciprocal_part}
\vec{e}_{\ast}\cdot\int_{|\hat{\vec{x}}'-\hat{\vec{x}}_{p}^{\prime}|=1} (-\vec{n})\cdot(-q_{1}\mathbb{I}+ \hat{\nabla}'\vec{v}_{1}+{\hat{\nabla}'\vec{v}_{1}}^{\intercal}) \,d\hat{S}' \\
=-\int_{\hat{\mathcal{F}}'} \hat{\vec{u}}_{\ast}\cdot\left(\hat{\Theta}(\vec{e}_{z'}\times\vec{v}_{0}) + \vec{v}_{0}\cdot\hat{\nabla}'\hat{\bar{\vec{u}}}+ (\vec{v}_{0}+\hat{\bar{\vec{u}}}-\hat{\Theta}(\vec{e}_{z'}\times\hat{\vec{x}}'))\cdot\hat{\nabla}'\vec{v}_{0}\right) \,d\hat{V}' ,
\end{multline} 
for $\ast=x',y',z'$ where each velocity $\hat{\vec{u}}_{\ast}$, along with a corresponding pressure $\hat{p}_{\ast}$, solves
\begin{subequations}\label{eqn:drag_coef_pde}
\begin{align}
-\hat{\nabla}'\hat{p}_{\ast}+\hat{\nabla}^{\prime\,2}\hat{\vec{u}}_{\ast}&=\vec{0} &&\text{on $\hat{\vec{x}}'\in\hat{\mathcal{F}}'$} , & \hat{\nabla}'\cdot\hat{\vec{u}}_{\ast}&=0  && \text{on $\hat{\vec{x}}'\in\hat{\mathcal{F}}'$} , \\
\hat{\vec{u}}_{\ast}&=\vec{0} &&\text{on $\hat{\vec{x}}'\in\partial\hat{\mathcal{D}}'$} , &
\hat{\vec{u}}_{\ast}&=\vec{e}_{\ast} &&\text{on $\hat{\vec{x}}'\in\partial\hat{\mathcal{F}}'\backslash\partial\hat{\mathcal{D}}'$} .
\end{align}
\end{subequations}
Note that we only need to consider the $x',z'$ components since the $y'$ component, which perturbs the particle motion in the $y'$ direction at $O(1)$, has no effect on the cross-section forces at the same order of magnitude. 
This approach cuts down on computation time (since each $\hat{p}_{\ast},\hat{\vec{u}}_{\ast}$ is computed to estimate drag coefficients, see Section \ref{sec:trajectory_model}), and also provides opportunities to examine how different components of the flow influence $\vec{F}_{0}$ in more detail (as discussed in Section~\ref{sec:as_separation}).
Additionally, note that $\vec{T}_{0}$ can be ignored since perturbations of the particle spin have no influence on the (linear) forcing to this order of approximation.
Given the boundary condition \eqref{eqn:perturb_eqn0_d}, $\vec{F}_{-1},\vec{T}_{-1}$ include drag and torque components that come from the secondary fluid motion.
In order to separate the effect of the axial and secondary flow components and better understand how each influences the particle motion it is convenient to break up $\hat{\bar{\vec{u}}},q_{0},\vec{v}_{0}$ and \eqref{eqn:perturb_eqn0} into two distinct parts.
This will be done in the following section.

\subsection{Separation of axial and secondary flow effects}\label{sec:as_separation}

Let 
\begin{equation}\label{eqn:bg_as_expansion}
\hat{\bar{\vec{u}}} = \hat{\bar{\vec{u}}}_{a} + \hat{\bar{\vec{u}}}_{s} ,
\end{equation}
where
\begin{subequations}
\begin{align*}
\hat{\bar{\vec{u}}}_{a} &:= \hat{\bar{u}}_{\theta}(r',z')\hat{\vec{e}}_{\theta'} 
= \frac{1}{U_{m}a/\ell}\bar{u}_{\theta}(r',z')\hat{\vec{e}}_{\theta'} , \\
\hat{\bar{\vec{u}}}_{s} &:= \hat{\bar{u}}_{r}(r',z')\hat{\vec{e}}_{r'}+\hat{\bar{u}}_{z}(r',z')\hat{\vec{e}}_{z'} 
= \frac{1}{U_{m}a/\ell}\bigg(\bar{u}_{r}(r',z')\hat{\vec{e}}_{r'}+\bar{u}_{z}(r',z')\hat{\vec{e}}_{z'}\bigg) ,
\end{align*}
\end{subequations}
denote the axial and secondary components respectively.
Similarly, we separate $q_{0},\vec{v}_{0}$ into two parts, specifically
\begin{equation}\label{eqn:vq_as_expansion}
\vec{v}_{0}=\vec{v}_{0,a}+\vec{v}_{0,s} , \qquad q_{0}=q_{0,a}+q_{0,s} , 
\end{equation}
such that the pairs $q_{0,a},\vec{v}_{0,a}$ and $q_{0,s},\vec{v}_{0,s}$ each solve \eqref{eqn:perturb_eqn0} except with the boundary condition \eqref{eqn:perturb_eqn0_d} replaced by
\begin{subequations}\label{eqn:perturb_eqn0_d_breakdown}
\begin{align}
\vec{v}_{0,a} &= \hat{\vec{\Omega}}_{p}^{\prime}\times(\hat{\vec{x}}'-\hat{\vec{x}}_{p}^{\prime}) +  \hat{\Theta}(\vec{e}_{z'}\times\hat{\vec{x}}^{\prime}) -\hat{\bar{\vec{u}}}_{a} , \quad &&\text{on $\hat{\vec{x}}'\in\partial\hat{\mathcal{F}}'\backslash\partial\hat{\mathcal{D}}'$} ,  \label{eqn:perturb_eqn0a_d} \\
\vec{v}_{0,s} &= -\hat{\bar{\vec{u}}}_{s} , \quad &&\text{on $\hat{\vec{x}}'\in\partial\hat{\mathcal{F}}'\backslash\partial\hat{\mathcal{D}}'$} , \label{eqn:perturb_eqn0s_d}
\end{align}
\end{subequations}
respectively.
The force and torque is similarly separated as $\vec{F}_{-1}=\vec{F}_{-1,a}+\vec{F}_{-1,s}$ and $\vec{T}_{-1}=\vec{T}_{-1,a}+\vec{T}_{-1,s}$. 
Specifically, for each $\ast=a,s$ one has
\begin{subequations}
\begin{align}
\vec{F}_{-1,\ast} &=
\int_{|\hat{\vec{x}}'-\hat{\vec{x}}_{p}^{\prime}|=1} (-\vec{n})\cdot\left(-q_{0,\ast}\mathbb{I}+ \hat{\nabla}'\vec{v}_{0,\ast}+{\hat{\nabla}'\vec{v}_{0,\ast}}^{\intercal}\right) \,d\hat{S}'  , \label{eqn:Fm1_ast} \\
\vec{T}_{-1,\ast} &= \int_{|\hat{\vec{x}}'-\hat{\vec{x}}_{p}^{\prime}|=1} (\hat{\vec{x}}'-\hat{\vec{x}}_{p}^{\prime})\times\left((-\vec{n})\cdot\left(-q_{0,\ast}\mathbb{I}+ \hat{\nabla}'\vec{v}_{0,\ast}+{\hat{\nabla}'\vec{v}_{0,\ast}}^{\intercal}\right)\right) \,d\hat{S}' . \label{eqn:Tm1_ast}
\end{align}
\end{subequations}

This separation of axial and secondary flow effects yields a clear methodology for solving the leading order disturbance equations.
One first solves for $q_{0,a},\vec{v}_{0,a}$ and in doing so determine $\hat{\Theta},\hat{\vec{\Omega}}_{p}^{\prime}$ such that $\vec{F}_{-1,a}=\vec{T}_{-1,a}=\vec{0}$.
The $q_{0,s},\vec{v}_{0,s}$ components are then solved independently from which $\vec{F}_{-1,s},\vec{T}_{-1,s}$ are calculated to provide the drag and torque on the particle from the secondary fluid motion.
The drag term $\vec{F}_{-1,s}$ can then be added to $\vec{F}_{0}$ to obtain the net perturbing forces on the particle (up to $O(1)$). 
While this drag from the secondary flow motion features as an $O(\Rep^{-1})$ factor it is typical that the magnitude of $\hat{\bar{\vec{u}}}_{s}$ is sufficiently small that the resulting force is comparable to $\vec{F}_{0}$ (since if this was not the case it is unlikely that focusing of particles would be observed since they would instead follow the streamlines of the secondary flow).

The $\vec{F}_{0}$ component of the force can also be expanded based on the separation of axial and secondary background flow components.
The volume integral over $\hat{\bar{\vec{u}}}\cdot\hat{\nabla}'\hat{\bar{\vec{u}}}$ in \eqref{eqn:F0} can be expanded as 
\begin{multline}\label{eqn:bg_inertia_as_expansion}
\int_{|\hat{\vec{x}}'-\hat{\vec{x}}_{p}^{\prime}|<1} \hat{\bar{\vec{u}}}_{a}\cdot\hat{\nabla}'\hat{\bar{\vec{u}}}_{a} \,d\hat{V}'
+\int_{|\hat{\vec{x}}'-\hat{\vec{x}}_{p}^{\prime}|<1} \hat{\bar{\vec{u}}}_{s}\cdot\hat{\nabla}'\hat{\bar{\vec{u}}}_{a}+\hat{\bar{\vec{u}}}_{a}\cdot\hat{\nabla}'\hat{\bar{\vec{u}}}_{s} \,d\hat{V}' \\
+\int_{|\hat{\vec{x}}'-\hat{\vec{x}}_{p}^{\prime}|<1} \hat{\bar{\vec{u}}}_{s}\cdot\hat{\nabla}'\hat{\bar{\vec{u}}}_{s} \,d\hat{V}' ,
\end{multline}
where $y',z'$ components of the first integral are exactly zero (and the $x'$ component is comparable to $(4\pi/3)\hat{\Theta}^{2}(\vec{e}_{z'}\times(\vec{e}_{z'}\times\hat{\vec{x}}_{p}^{\prime}))$ for a small neutrally buoyant particle), the second integral consists of only  a small $y'$ component
and the third integral provides a small force in the $x',z'$ directions. 
Utilising \eqref{eqn:F0_reciprocal_part}, the $\ast=x',z'$ components of the surface integral over the fluid stress from $q_{1},\vec{v}_{1}$ in \eqref{eqn:F0} can each be expanded as
\begin{subequations}\label{eqn:F0_reciprocal_part_as_expansion}
\begin{align}
&-\int_{\hat{\mathcal{F}}'} \hat{\vec{u}}_{\ast}\cdot\left(\hat{\Theta}\vec{e}_{z'}\times\vec{v}_{0,a} + \vec{v}_{0,a}\cdot\hat{\nabla}'\hat{\bar{\vec{u}}}_{a}+ (\vec{v}_{0,a}+\hat{\bar{\vec{u}}}_{a}-\hat{\Theta}\vec{e}_{z'}\times\hat{\vec{x}}')\cdot\hat{\nabla}'\vec{v}_{0,a}\right) \,d\hat{V}' \\
&-\int_{\hat{\mathcal{F}}'} \hat{\vec{u}}_{\ast}\cdot\left( \vec{v}_{0,s}\cdot\hat{\nabla}'\hat{\bar{\vec{u}}}_{s}+ (\vec{v}_{0,s}+\hat{\bar{\vec{u}}}_{s})\cdot\hat{\nabla}'\vec{v}_{0,s}\right) \,d\hat{V}' \\
&-\int_{\hat{\mathcal{F}}'} \hat{\vec{u}}_{\ast}\cdot\left(\hat{\Theta}\vec{e}_{z'}\times\vec{v}_{0,s} - (\hat{\Theta}\vec{e}_{z'}\times\hat{\vec{x}}')\cdot\hat{\nabla}'\vec{v}_{0,s} +\vec{v}_{0,a}\cdot\hat{\nabla}'\hat{\bar{\vec{u}}}_{s} \right. \\
&\qquad\qquad\quad\left. +\vec{v}_{0,s}\cdot\hat{\nabla}'\hat{\bar{\vec{u}}}_{a}+(\vec{v}_{0,a}+\hat{\bar{\vec{u}}}_{a})\cdot\hat{\nabla}'\vec{v}_{0,s}+ (\vec{v}_{0,s}+\hat{\bar{\vec{u}}}_{s})\cdot\hat{\nabla}'\vec{v}_{0,a}\right) \,d\hat{V}' . \notag
\end{align}
\end{subequations}
Here the first and second integrals provide a force in the $x',z'$ directions and the third integral provides a small force in the $y'$ directions.
Note that, in the context of understanding the particle behaviour within the cross-section, the $y'$ components of $\vec{F}_{0}$ can essentially be ignored since the small perturbation to the axial motion has no further influence on the cross-section forcing to this order of magnitude.
As such, one can discard the second and third component of \eqref{eqn:bg_inertia_as_expansion} and \eqref{eqn:F0_reciprocal_part_as_expansion} respectively.

In summary, the force on the particle within the cross-section up to $O(1)$ is given by
\begin{equation}\label{eqn:force_perturb}
\vec{F}_{p}^{\prime} := \big(\vec{e}_{x'}\cdot(\Rep^{-1}\vec{F}_{-1,s}+\vec{F}_{0})\big)\vec{e}_{x'} + \big(\vec{e}_{z'}\cdot(\Rep^{-1}\vec{F}_{-1,s}+\vec{F}_{0})\big)\vec{e}_{z'} .
\end{equation}
Given a fixed particle size and duct size/shape, then $\vec{F}_{p}^{\prime}$ can be considered as a function of the particle position within the cross-section (i.e. $\hat{x}_{p,r'},\hat{x}_{p,z'}$, recalling that at each point each of $\hat{\Theta},\hat{\vec{\Omega}}_{p}^{\prime}$ are chosen such that such that $\vec{F}_{-1,a}=\vec{T}_{-1,a}=\vec{0}$).

\subsection{A first order model of particle trajectories}\label{sec:trajectory_model}

%
Here we consider a simple model for the trajectory of a particle within the cross-section based on the terminal velocity, that is we take the particle velocity in the $r',z'$ directions to be such that the drag force is equal and opposite to the perturbing force $\vec{F}_{p}^{\prime}(\hat{x}_{p,r'},\hat{x}_{p,z'})$.
In order to use this model we must first determine the drag coefficients for a particle with non-zero velocity in the cross-sectional plane.

Suppose the spherical particle is moving with a velocity $\hat{u}_{p,r'}^{\prime},\hat{u}_{p,z'}^{\prime}$ in the $r',z'$ directions respectively, then the cross-sectional drag force on the particle (non-dimensionalised with respect to the force scale $\rho U_{m}^{2}a^{4}/\ell^{2}$) can be expressed as $\Rep^{-1}(C_{r'}\hat{u}_{p,r'}\hat{\vec{e}}_{r'}+C_{z'}\hat{u}_{p,z'}\hat{\vec{e}}_{z'})$ where $C_{r'},C_{z'}$ are (dimensionless) drag coefficients. 
For an asymptotically small particle away from the walls the (dimensionless) drag coefficients could be estimated via Stokes drag law as simply $6\pi$. 
However, in the context of a microfluidic duct, the finite particle size and proximity of the duct walls have an effect such that the true drag coefficients are noticeably larger than this.
A better estimate of the drag coefficients is obtained by explicitly computing the drag force on the particle.
Note that the $\hat{p}_{\ast},\hat{\vec{u}}_{\ast}$ from Section~\ref{sec:Rep_expansion} which each solve \eqref{eqn:drag_coef_pde} describe the Stokes flow resulting from a particle moving through a stationary fluid in the duct with velocity $\vec{e}_{\ast}$. 
Thus, we can re-use these solutions to compute the drag coefficients.
Specifically, for each $C_{\ast}$ we need only integrate the resulting fluid stress over the surface of the particle, that is
\begin{equation*}
C_{\ast} = \hat{\vec{e}}_{\ast}\cdot\int_{|\hat{\vec{x}}'-\hat{\vec{x}}_{p}^{\prime}|=1} (-\vec{n})\cdot(-\hat{p}_{\ast}\mathbb{I}+ \hat{\nabla}'\hat{\vec{u}}_{\ast}+{\hat{\nabla}'\hat{\vec{u}}_{\ast}}^{\intercal}) \,d\hat{S}'  .
\end{equation*}
Similar to $\vec{F}_{p}^{\prime}$, given a fixed particle size and duct size/shape the drag coefficients $C_{r'},C_{z'}$ depend primarily on the particle position within the cross-section.

Therefore, with the addition of these drag terms, the cross-sectional force on the particle is
\begin{equation}
\left(\Rep^{-1}C_{r'}\hat{u}_{p,r'}^{\prime}+\hat{F}_{p,r'}^{\prime}\right)\hat{\vec{e}}_{r'}+\left(\Rep^{-1}C_{z'}\hat{u}_{p,z'}^{\prime}+\hat{F}_{p,z'}^{\prime}\right)\hat{\vec{e}}_{z'} .
\end{equation}
Thus, equating this to zero and writing this out in terms of $\hat{x}_{p,r'}^{\prime},\hat{x}_{p,z'}^{\prime}$, we obtain the first order system of ordinary differential equations
\begin{equation}\label{eqn:cs_traj_model}
\frac{d \hat{x}_{p,r'}^{\prime}}{d t'} = -\Rep\frac{\hat{F}_{p,r'}^{\prime}(\hat{x}_{p,r'}^{\prime},\hat{x}_{p,z'}^{\prime})}{C_{r'}(\hat{x}_{p,r'}^{\prime},\hat{x}_{p,z'}^{\prime})} , \quad
\frac{d \hat{x}_{p,z'}^{\prime}}{d t'} = -\Rep\frac{\hat{F}_{p,z'}^{\prime}(\hat{x}_{p,r'}^{\prime},\hat{x}_{p,z'}^{\prime})}{C_{z'}(\hat{x}_{p,r'}^{\prime},\hat{x}_{p,z'}^{\prime})} .
\end{equation}
The solution of these coupled equations approximates the particle trajectory within the cross-section.
Interpreting $\hat{\Theta}$ as a function of $\hat{x}_{p,r'}^{\prime},\hat{x}_{p,z'}^{\prime}$ (recalling that $\hat{\Theta},\hat{\vec{\Omega}}_{p}^{\prime}$ are chosen at each $(\hat{x}_{p,r'}^{\prime},\hat{x}_{p,z'}^{\prime})$ such that $\vec{F}_{-1,a}=\vec{0}$), the angular position of the particle can also be estimated via $d \theta_{p}/d t' = \hat{\Theta}(\hat{x}_{p,r'}^{\prime},\hat{x}_{p,z'}^{\prime})$.

\section{Estimation of particle behaviour at low flow rates}\label{sec:lfr_limit}

Experiments involving curved and spiral ducts have been done with $U_{m},\Re$ as high as $O(1),O(100)$ respectively (in order to achieve reasonable throughput/flow-rate).  
That said, the low flow rate case it is particularly interesting for several reasons. 
First, the behaviour at low flow rate in a curved duct differs from that in a straight duct, i.e. the behaviour as $U_{m}\rightarrow0$ (or equivalently $\Re\rightarrow 0$) differs from that as $R\rightarrow\infty$. 
As such, understanding what happens at low flow rates is a good first step towards understanding the difference between straight and curved ducts.
Second, it allows us to reduce parameter space since the focusing behaviour becomes independent of flow rate; we are able to thoroughly examine how focusing behaviour is influenced by the bend radius and particle size in this regime.
Third, we can additionally consider the limit $R\rightarrow\infty$ to check that the focusing positions in a straight duct are recovered in order to validate the model.
Lastly, the assumption is not as restrictive as it may first seem and we expect the predictions may be applicable up to at least $\Re=O(10)$.

The scaling employed to study the inertial lift force differs from the characteristic scales in the background flow; we return to the physical background flow velocity $\bar{\vec{u}}$ to examine the scale of its components.
Recalling that $U_{m}$ denotes the maximum of $\bar{u}_{\theta}(r,z)$, and letting $\epsilon:=\ell/(2R)$, then it can be shown that if $K:=\epsilon\Re^{2}/4$ is sufficiently small then
\begin{equation}\label{eqn:secondary_flow_scale}
\bar{u}_{r}(r,z),\bar{u}_{z}(r,z) \propto (\epsilon \Re/2) U_{m} .
\end{equation}
This scaling is discussed \citet{Harding2018} in which the governing equations \eqref{eqn:ns_bg} are explicitly non-dimensionalised (a brief account is also provided in Appendix~\ref{app:bgf} for completeness).
Furthermore, the dimensionless velocity components can be expressed in terms of a perturbation expansion in $K$, specifically
\begin{equation}\label{eqn:background_flow_expansion}
\left(\frac{\bar{u}_{\theta}}{U_{m}},\frac{\bar{u}_{r}}{(\epsilon\Re/2) U_{m}},\frac{\bar{u}_{z}}{(\epsilon\Re/2) U_{m}}\right)=\sum_{i=0}^{\infty}K^{i}(\bar{u}_{\theta,i},\bar{u}_{r,i},\bar{u}_{z,i}) .
\end{equation}
This series leads to a linearisation of the governing equations in which each $\bar{u}_{\theta,i}$ and pair $\bar{u}_{r,i},\bar{u}_{z,i}$ (which are combined into a stream function $\Phi_{i}$) can be solved in an alternating fashion.

It has been demonstrated that the terms in the expansion \eqref{eqn:background_flow_expansion} decay in a manner such that the expansion generally converges whenever $K\lessapprox 200$~\citep{Harding2018}.
Another observation that can be made from the results of \citet{Harding2018} is that the $i=0$ terms provide a good approximation for each velocity component (e.g. within a few percent) whenever $\Re^{2}\lessapprox 10/\epsilon$.
Since $\epsilon\ll10^{-1}$ is typical in spiral microfluidic ducts one can generally expect the $i=0$ terms to provide a reasonable approximation up to at least $\Re=O(10)$.
Furthermore, although \eqref{eqn:secondary_flow_scale} implies the secondary flow scales with $U_{m}^{2}$, and therefore vanishes quickly in the limit of small flow rate, it is important to note that the inertial lift force also scales with $U_{m}^{2}$.
A consequence of this is the $i=0$ terms in the expansion of the background flow are sufficient to approximate the cross-sectional force on the particle when the flow rate is small.
To make this clear we further examine the expansion of $\vec{F}_{p}^{\prime}$ into its different parts depending on the axial and secondary components of the background flow.

Returning to the non-dimensional setting from Section~\ref{sec:analysis} we have
\begin{subequations}\label{eqn:bg_as_approx}
\begin{align}
\hat{\bar{\vec{u}}}_{a}&= \frac{U_{m}\bar{u}_{\theta}\vec{e}_{\theta'}}{U_{m}a/\ell}=\frac{\ell}{a}\bar{u}_{\theta}\vec{e}_{\theta'} , \\
\hat{\bar{\vec{u}}}_{s}&= \frac{(\epsilon\Re/2) U_{m}}{U_{m}a/\ell}(\bar{u}_{r}\vec{e}_{r'}+\bar{u}_{z}\vec{e}_{z'})=\frac{\ell^{2}\Re}{4aR}(\bar{u}_{r}\vec{e}_{r'}+\bar{u}_{z}\vec{e}_{z'}) .
\end{align}
\end{subequations}
Given it is now clear that $\hat{\bar{\vec{u}}}_{s}\propto \ell^{2}\Re/4aR$ it follows that $\vec{v}_{0,s}\propto\ell^{2}\Re/4aR$, and consequently
$$
\Rep^{-1}\vec{F}_{-1,s} \propto \Rep^{-1}\frac{\ell^{2}\Re}{4aR} = \frac{\ell^{4}}{4a^{3}R} .
$$
Thus the contribution of the secondary flow drag to $\vec{F}_{p}^{\prime}$ scales with the dimensionless constant $\kappa:=\ell^{4}/4a^{3}R$ which depends only on physical length scales.
At the same time it can be seen that the $\vec{F}_{0}$ contribution does not change in magnitude with respect to physical lengths.
In particular, note that $\vec{F}_{0}$ can be approximated as
\begin{align*}
\vec{F}_{0}&\approx -\frac{\rho_{p}-\rho}{\rho}\frac{4\pi}{3}\hat{g}\vec{e}_{z}
-\frac{\rho_{p}}{\rho}\frac{4\pi}{3} \hat{\Theta}^{2}(\vec{e}_{z'}\times(\vec{e}_{z}\times\hat{\vec{x}}_{p}^{\prime})) 
+\int_{|\hat{\vec{x}}'-\hat{\vec{x}}_{p}^{\prime}|<1} \hat{\bar{\vec{u}}}_{a}\cdot\hat{\nabla}'\hat{\bar{\vec{u}}}_{a} \,d\hat{V}' \notag \\
&\quad-\vec{e}_{x'}\int_{\hat{\mathcal{F}}'} \hat{\vec{u}}_{x'}\cdot\left(\hat{\Theta}\vec{e}_{z}\times\vec{v}_{0,a} + \vec{v}_{0,a}\cdot\hat{\nabla}'\hat{\bar{\vec{u}}}_{a}+ (\vec{v}_{0,a}+\hat{\bar{\vec{u}}}_{a}-\hat{\Theta}\vec{e}_{z}\times\hat{\vec{x}}')\cdot\hat{\nabla}'\vec{v}_{0,a}\right) \,d\hat{V}' \\
&\quad-\vec{e}_{z'}\int_{\hat{\mathcal{F}}'} \hat{\vec{u}}_{z'}\cdot\left(\hat{\Theta}\vec{e}_{z}\times\vec{v}_{0,a} + \vec{v}_{0,a}\cdot\hat{\nabla}'\hat{\bar{\vec{u}}}_{a}+ (\vec{v}_{0,a}+\hat{\bar{\vec{u}}}_{a}-\hat{\Theta}\vec{e}_{z}\times\hat{\vec{x}}')\cdot\hat{\nabla}'\vec{v}_{0,a}\right) \,d\hat{V}' ,
\end{align*}
since the additional terms in \eqref{eqn:bg_inertia_as_expansion} and \eqref{eqn:F0_reciprocal_part_as_expansion} involving $\hat{\bar{\vec{u}}}_{s}$ vanish much faster than the terms involving $\hat{\bar{\vec{u}}}_{a}$, which we have retained.
Hence, the relative magnitude of the secondary flow drag $\Rep^{-1}\vec{F}_{-1,s}$ relative to $\vec{F}_{0}$ is $\kappa=\ell^{4}/4a^{3}R$ is consistent with a simplified study of the forces on the particle in the limit of large $R$ and small flow rate~\citep{Harding2018_2}.
Furthermore, it follows that we need only a leading order approximation of $\hat{\bar{\vec{u}}}_{s}$ to estimate $\vec{F}_{-1,s}$ and a leading order approximation of $\hat{\bar{\vec{u}}}_{a}$ to estimate $\vec{F}_{0}$.
That is, it suffices to take
\begin{equation*}
\hat{\bar{\vec{u}}}_{a}\approx\frac{\ell}{a}\bar{u}_{\theta,0}\vec{e}_{\theta'}  
\qquad\text{and}\qquad
\hat{\bar{\vec{u}}}_{s}\approx\frac{\ell^{2}\Re}{4aR}(\bar{u}_{r,0}\vec{e}_{r'}+\bar{u}_{z,0}\vec{e}_{z'}) .
\end{equation*}

The scaling of $\Rep^{-1}\vec{F}_{-1,s}$ with $\kappa$ makes it clear that, as $\epsilon=\ell/(2R)\rightarrow 0$, the relative effect of the secondary flow drag becomes insignificant leaving only $\vec{F}_{0}$.
Furthermore, $\vec{F}_{0}$ approaches the inertial lift force obtained in a straight duct since
\begin{align*}
\lim_{\epsilon\rightarrow0}\vec{F}_{0} &=-\vec{e}_{r'} \int_{\hat{\mathcal{F}}'} \hat{\vec{u}}_{r'}\cdot\left(\vec{v}_{0,a}\cdot\hat{\nabla}'\hat{\bar{\vec{u}}}_{a}+ (\vec{v}_{0,a}+\hat{\bar{\vec{u}}}_{a}-U_{p}\vec{e}_{y'})\cdot\hat{\nabla}'\vec{v}_{0,a}\right) \,d\hat{V}' \\
&\quad-\vec{e}_{z'} \int_{\hat{\mathcal{F}}'} \hat{\vec{u}}_{z'}\cdot\left(\vec{v}_{0,a}\cdot\hat{\nabla}'\hat{\bar{\vec{u}}}_{a}+ (\vec{v}_{0,a}+\hat{\bar{\vec{u}}}_{a}-U_{p}\vec{e}_{y'})\cdot\hat{\nabla}'\vec{v}_{0,a}\right) \,d\hat{V}' ,
\end{align*}
where $\hat{\bar{\vec{u}}}_{a}$ converges to a classical Poiseuille flow profile and $\vec{v}_{0,a}$ converges to the appropriate disturbance flow. 
On the other hand, with fixed physical dimensions (and thus fixed $\epsilon$) the relative size of the secondary drag force in relation to the other forces making up $\vec{F}_{p}^{\prime}$ remains constant.
This implies that in the limit of small flow rate the interaction between the inertial lift force and secondary flow drag force converge to something which is different than that in the case of a straight duct.
In Section~\ref{sec:results} we examine this particular limit in detail for several cross-sections, particle sizes and bend radii.

\section{Computational results and discussion}\label{sec:results}

A finite element method is used to solve each Stokes problem, implemented using the FEniCS software \citep{LoggMardalEtAl2012a,AlnaesBlechta2015a}.
The computational domain consists of only a portion of the curved duct, in a neighbourhood of the particle, discretised by a mesh consisting of $O(10^{6})$ tetrahedra refined such that those elements near the particle have much smaller scale than those far away (since the disturbance varies most near the particle).
Continuous Taylor--Hood elements are used, specifically a quadratic and linear polynomial basis for the velocity and pressure respectively.
The Dirichlet boundary conditions on the duct walls and particle surface are enforced explicitly (i.e. at the linear algebraic level) and natural (stress-free) boundary conditions are applied at the ends of the (truncated) curved duct.
The approximation of background flow terms $\hat{\bar{\vec{u}}}_{a},\hat{\bar{\vec{u}}}_{s}$ as in \eqref{eqn:bg_as_approx} are precomputed using the Rayleigh--Ritz method described in \citet{Harding2018} and subsequently interpolated in the finite element space. 
With $\vec{F}_{p}^{\prime}$ and drag coefficients $C_{r'},C_{z'}$ estimated at a large number of (equidistant) points in the cross-section we then construct a bivariate cubic spline to interpolate the data.
The smooth interpolants are then used for subsequent analysis, for example the estimation of particle trajectories via the solution of \eqref{eqn:cs_traj_model}.
The behaviour of particles suspended in flow through curved ducts having square, rectangular and trapezoidal shaped cross-sections is examined.
We use the approximation for low flow rate developed in Section~\ref{sec:lfr_limit} and particles are assumed to be neutrally buoyant (i.e. $\rho_{p}=\rho$).

\subsection{Ducts having a square cross-section}

\begin{figure}
\begin{center}
\begin{subfigure}[b]{0.48\textwidth}
\centering
\includegraphics{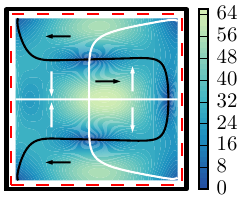}
\caption{$a=0.05$}
\end{subfigure}
\begin{subfigure}[b]{0.48\textwidth}
\centering
\includegraphics{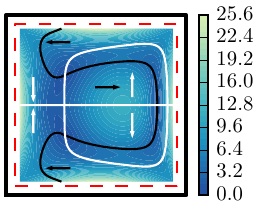}
\caption{$a=0.10$}
\end{subfigure}
\\
\begin{subfigure}[b]{0.48\textwidth}
\centering
\includegraphics{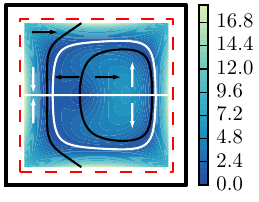}
\caption{$a=0.15$}
\end{subfigure}
\begin{subfigure}[b]{0.48\textwidth}
\centering
\includegraphics{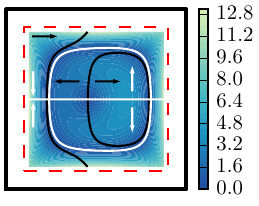}
\caption{$a=0.20$}
\end{subfigure}
\end{center}
\caption{
The force $\vec{F}_{p}^{\prime}$ on neutrally buoyant particles at different locations in a curved duct with square cross-section ($W=H=2$) and bend radius $R=160$.
The colour background shows the magnitude of $\vec{F}_{p}^{\prime}$. 
Black and white contours are the zero level set curves of the horizontal and vertical components respectively whereas the arrows indicate the sign of each component in the area bounded by the respective contour.
The left wall is on the inside of the bend.
The dashed red line shows where the centre of the particle lies when its surface touches the wall.
}\label{fig:square_lift_plots_R160}
\end{figure}

\begin{figure}
\begin{center}
\begin{subfigure}[b]{0.244\textwidth}
\centering
\includegraphics{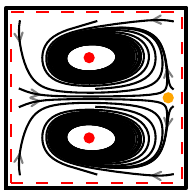}
\caption{$a=0.05$}
\end{subfigure}
\begin{subfigure}[b]{0.244\textwidth}
\centering
\includegraphics{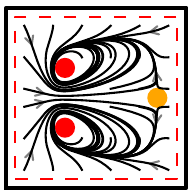}
\caption{$a=0.10$}
\end{subfigure}
\begin{subfigure}[b]{0.244\textwidth}
\centering
\includegraphics{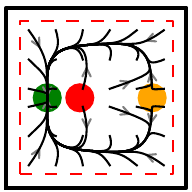}
\caption{$a=0.15$}
\end{subfigure}
\begin{subfigure}[b]{0.244\textwidth}
\centering
\includegraphics{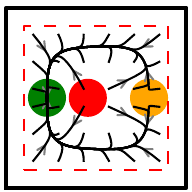}
\caption{$a=0.20$}
\end{subfigure}
\end{center}
\caption{
Approximate trajectories of neutrally buoyant particles within a curved duct with square cross-section ($W=H=2$) and bend radius $R=160$.
Trajectories from several starting positions have been super-imposed.
Green, orange and red markers show the location of stable, (unstable) saddle and unstable equilibria respectively (with marker size indicative of particle size).
The left wall is on the inside of the bend.
The dashed red line shows where the centre of the particle lies when its surface touches the wall.
}\label{fig:square_traj_plots_R160}
\end{figure}

For a duct having a square cross-section with $W=H=2$ the cross-sectional force and drag on particles is computed for the particle sizes $a\in\{0.20,0.15,0.10,0.05\}$ and bend radii $R\in\{640,320,160,80,40\}$ (or equivalently $\epsilon^{-1}\in\{640,320,160,80,40\}$).
Figures~\ref{fig:square_lift_plots_R160} and \ref{fig:square_traj_plots_R160} show the force and trajectory plots respectively for the specific bend radius $R=160$.
Results for the smallest particle ($a=0.05$) are depicted in Figures~\ref{fig:square_lift_plots_R160}(a) and \ref{fig:square_traj_plots_R160}(a).
Here $\vec{F}_{p}^{\prime}$ is dominated by the secondary flow drag since $\ell^{4}/4a^{3}R=200$ is reasonably large.
Three equilibria can be identified in the force plot, one near the right (outer) wall and two which are near the centre horizontally and vertically symmetric about the $r$ axis.
The equilibrium near the outer wall is an unstable saddle (it attracts horizontally but repels vertically). 
The remaining two are difficult to infer visually from the force plot but are also unstable (both eigenvalues of the Jacobian have positive real part, albeit quite small).
The trajectory plot makes it clear that particles are strongly affected by the vortices of the secondary flow.
It is notable the particles starting too close to the walls migrate inwards a little before becoming trapped in the vortex motion.

Results for the second smallest particle ($a=0.10$ and $\ell^{4}/4a^{3}R=25$) are depicted in Figures~\ref{fig:square_lift_plots_R160}(b) and \ref{fig:square_traj_plots_R160}(b).
Three equilibria can again be identified from the force plot, each of which is unstable, similar to those in (a).
Notice that the two symmetric equilibria are now closer to the inside wall.
The trajectory plot again shows that the vortex motion of the secondary flow has a strong influence on the trajectories, however, in this case the particles appear to migrate onto one of two (symmetrical) stable orbits relatively quickly.
This suggests the inertial lift force and secondary flow drag are similar in magnitude.

Results for the second largest particle ($a=0.15$ and $\ell^{4}/4a^{3}R\approx7.41$) are depicted in Figures~\ref{fig:square_lift_plots_R160}(c) and \ref{fig:square_traj_plots_R160}(c). 
Three equilibria can be identified from the force plot, each centred vertically within the cross-section. one near the centre of the outside wall and another near the centre of the inside wall.
The equilibrium near the inside wall is stable while the remaining two are unstable (the one nearest to the outside being a saddle point).
The trajectory plot illustrates that inertial lift force is the dominant effect  with complete focusing onto the stable equilibrium.
Particles initially migrate onto a `slow manifold' along which they more slowly migrate towards the unique stable equilibrium.

Results for the largest particle ($a=0.20$ and $\ell^{4}/4a^{3}R=3.125$) are depicted in Figures~\ref{fig:square_lift_plots_R160}(d) and \ref{fig:square_traj_plots_R160}(d) and are similar to those in Figures~\ref{fig:square_lift_plots_R160}(c) and \ref{fig:square_traj_plots_R160}(c).
The trajectory plot shows a more direct convergence onto the slow manifold, given the weaker effect of the secondary flow drag in this case.
Migration along the slow manifold is quite slow, as evidenced by the close proximity of the black and white contours in the force plot.
Note that for a particle which is a little larger, and/or a larger bend radius, we expect these contours to cross-over leading to additional equilibria. 
In particular, a stable equilibria may form near the centre of the outside wall, although relatively few particles would be expected to migrate towards it.

\begin{figure}
\begin{center}
\begin{subfigure}[b]{0.244\textwidth}
\centering
\includegraphics{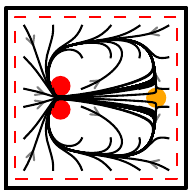}
\caption{$a=0.10$, $R=210$}
\end{subfigure}
\begin{subfigure}[b]{0.244\textwidth}
\centering
\includegraphics{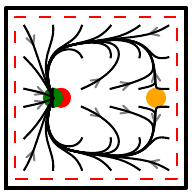}
\caption{$a=0.10$, $R=220$}
\end{subfigure}
\begin{subfigure}[b]{0.244\textwidth}
\centering
\includegraphics{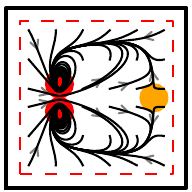}
\caption{$a=0.15$, $R=83$}
\end{subfigure}
\begin{subfigure}[b]{0.244\textwidth}
\centering
\includegraphics{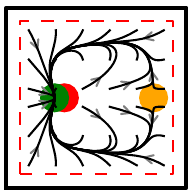}
\caption{$a=0.15$, $R=90$}
\end{subfigure}
\end{center}
\caption{
Approximate trajectories of neutrally buoyant particles within a curved duct with square cross-section ($W=H=2$).
Trajectories from several starting positions have been super-imposed.
Green, orange and red markers show the location of stable, (unstable) saddle and unstable equilibria respectively (with marker size indicative of particle size).
The left wall is on the inside of the bend.
The dashed red line shows where the centre of the particle lies when its surface touches the wall.
}\label{fig:square_traj_plots_Rvar}
\end{figure}

For other bend radii the results are qualitatively similar to those described above given a similar value of the ratio $\kappa=\ell^{4}/4a^{3}R$.
We briefly comment on some general trends.
For very large $\kappa$ particles effectively follow the vortex motion of the background flow (apart from some initial migration away from the walls).
In the limit $R\rightarrow\infty$ (or equivalently $\kappa\rightarrow0$ given fixed $a,\ell$) we recover the results for a straight duct, specifically the four stable equilibria near the centre of each wall as in \citet{HoodLeeRoper2015}. 
It is notable that $R\gg10^{3}$ is required before particle focusing similar to that in a straight duct is observed because even a very small influence from secondary flow motion is enough to modify the stability of some equilibria.
%

%
The transition from non-focusing to focusing behaviour is quite interesting.
As $R$ increases (and $\kappa$ decreases), the particles begin to focus towards trapping orbits as is the case observed for $a=0.10$ and $R=160$.
For the specific case of $a=0.10$, as $R$ continues to increase the size of the trapping orbits grows until they eventually meet and break up (around $\kappa\in[18,19]$) leaving behind the slow manifold and a stable equilibrium as depicted in Figures~\ref{fig:square_traj_plots_Rvar}(a,b).
Interestingly this is somewhat different for the slightly larger particle with $a=0.15$ in which the trapping orbits instead shrink while migrating towards what becomes the stable equilibrium (around $\kappa\in[13,14]$) as depicted in Figures~\ref{fig:square_traj_plots_Rvar}(c,d).
At this stage the significance of the trapping orbits and their structure are unclear and warrant further investigation.

Notice that the larger particles appear to focus near the inside wall while smaller ones become trapped in orbits a finite distance away from the inside wall. 
This provides a potential mechanism for separating larger particles from smaller ones in a fluid sample that contains a low concentration of both.
For example, by splitting the flow at the outlet into two streams which separate the inside $1/3$ and outside $2/3$ (measured laterally) one would expect the majority of larger particles to be collected from the inside stream (with a reduced portion of smaller particles).
However, the orbits onto which smaller particles are trapped are perhaps too close to the focusing location of the larger particles in a square duct for this to provide robust and efficient separation in practice.
Rectangular ducts small aspect ratio are more common in experiments that demonstrate separation, so we move onto examine two such cross-sections.

\subsection{Ducts having a rectangular cross-section}

\begin{figure}
\begin{center}
\begin{subfigure}[b]{0.48\textwidth}
\centering
\includegraphics{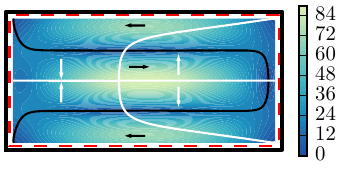}
\caption{$a=0.05$}
\end{subfigure}
\begin{subfigure}[b]{0.48\textwidth}
\centering
\includegraphics{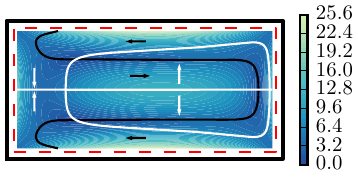}
\caption{$a=0.10$}
\end{subfigure}
\\
\begin{subfigure}[b]{0.48\textwidth}
\centering
\includegraphics{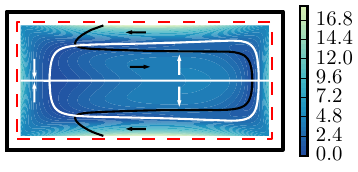}
\caption{$a=0.15$}
\end{subfigure}
\begin{subfigure}[b]{0.48\textwidth}
\centering
\includegraphics{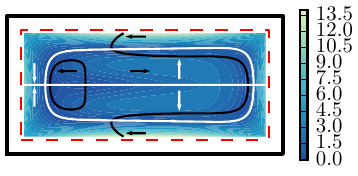}
\caption{$a=0.20$}
\end{subfigure}
\end{center}
\caption{
The force $\vec{F}_{p}^{\prime}$ on neutrally buoyant particles at different locations in a curved duct with rectangular cross-section ($W/2=H=2$) and bend radius $R=160$.
The colour background shows the magnitude of $\vec{F}_{p}^{\prime}$. 
Black and white contours are the zero level set curves of the horizontal and vertical components respectively whereas the arrows indicate the sign of each component in the area bounded by the respective contour.
The left wall is on the inside of the bend.
The dashed red line shows where the centre of the particle lies when its surface touches the wall.
}\label{fig:rect_lift_plots_R160}
\end{figure}

\begin{figure}
\begin{center}
\begin{subfigure}[b]{0.48\textwidth}
\centering
\includegraphics{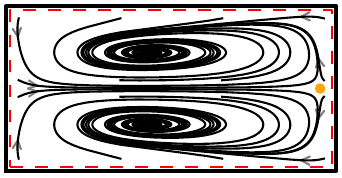}
\caption{$a=0.05$}
\end{subfigure}
\begin{subfigure}[b]{0.48\textwidth}
\centering
\includegraphics{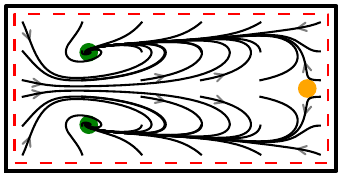}
\caption{$a=0.10$}
\end{subfigure}
\\
\begin{subfigure}[b]{0.48\textwidth}
\centering
\includegraphics{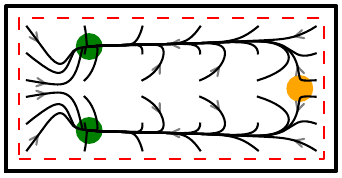}
\caption{$a=0.15$}
\end{subfigure}
\begin{subfigure}[b]{0.48\textwidth}
\centering
\includegraphics{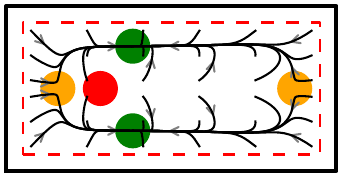}
\caption{$a=0.20$}
\end{subfigure}
\end{center}
\caption{
Approximate trajectories of neutrally buoyant particles in a curved duct with rectangular cross-section ($W/2=H=2$) and bend radius $R=160$.
Trajectories from several starting positions have been super-imposed.
Green, orange and red markers show the location of stable, (unstable) saddle and unstable equilibria respectively (with marker size indicative of particle size).
The left wall is on the inside of the bend.
The dashed red line shows where the centre of the particle lies when its surface touches the wall.
}\label{fig:rect_traj_plots_R160}
\end{figure}

Two rectangular cross-sections are considered, specifically with $H=2$ and $W\in\{4,8\}$.
For each we computed $\vec{F}_{p}^{\prime}$ for particles of size $a\in\{0.20,0.15,0.10,0.05\}$ and bend radii $R\in\{640,320,160,80\}$ (or equivalently $\epsilon^{-1}\in\{640,320,160,80\}$).
The force and trajectory plots for $W=4$ and $R=160$ are shown in Figures~\ref{fig:rect_lift_plots_R160} and \ref{fig:rect_traj_plots_R160} respectively.
Results for the smallest particle ($a=0.05$ and $\ell^{4}/4a^{3}R=200$) are depicted in Figures~\ref{fig:rect_lift_plots_R160}(a) and \ref{fig:rect_traj_plots_R160}(a).
As in the case of the square duct, the effect of secondary flow drag is the dominant cross-sectional force.
The force plot is similar to that in the square duct case for $a=0.05$ (only stretched width-wise) and, in particular, three equilibria can be identified.
The equilibrium near the right wall is again an unstable saddle.
However, unlike the square duct case, the remaining (symmetric) equilibria pair is stable in this case. 
This is evident in the trajectory plot where particles are observed to slowly converge towards the equilibria despite the relatively strong vortex motion. 
Note that the location of the stable pair is slightly left of the centre (the green marker is hidden by the spiral trajectories).
Thus, despite a strong influence from the secondary flow drag, the particles eventually becomes focused.

Results for the second smallest particle ($a=0.10$ and $\ell^{4}/4a^{3}R=25$) are depicted in Figures~\ref{fig:rect_lift_plots_R160}(b) and \ref{fig:rect_traj_plots_R160}(b). 
Again, the force plot could be interpreted as a stretched version of the square duct case.
The equilibrium near the right wall is an unstable saddle whereas the remaining (symmetric) pair is stable.
Notably the stable pair is much closer to the inside wall than in the case of the smallest particle.
The trajectory plot demonstrates that although the secondary flow drag and inertial lift force are expected to have similar magnitude, the particle converges onto a slow manifold before completing a full orbit around an equilibrium and then proceeds to migrate to the equilibrium.
This too is quite different from the square duct case despite the seemingly similar structure of the force plots.

Results for the second largest particle ($a=0.15$ and $\ell^{4}/4a^{3}R\approx7.41$) are depicted in Figures~\ref{fig:rect_lift_plots_R160}(c) and \ref{fig:rect_traj_plots_R160}(c).
Three equilibria can again be identified, the unstable saddle near the outside wall and a stable pair.
The stable pair is slightly closer to the centre than in the case of $a=0.10$, but still closer to the inside wall than the case $a=0.05$.
The trajectory plot shows particles converge onto the slow manifold with minimal impact from the secondary flow motion before then migrating towards the stable equilibria.

Results for the largest particle ($a=0.20$ and $\ell^{4}/4a^{3}R=3.125$) are depicted in Figures~\ref{fig:rect_lift_plots_R160}(d) and \ref{fig:rect_traj_plots_R160}(d).
Two new unstable equilibria can be identified in the force plot located on the horizontal symmetry line on the left side.
Additionally, the stable equilibria pair has shifted further towards the right.
Inertial lift force is the dominant effect in this case and the trajectory plot shows a more direct convergence onto the slow manifold prior to convergence towards the stable equilibria.

The results for different values of $R$ are similar to those described above given a similar value of $\kappa=\ell^{4}/4a^{3}R$ and so we remark on some general observations and trends.
In the limit $R\rightarrow\infty$ (equivalently $\kappa=0$ for fixed $a,\ell$) we recover the focusing behaviour of particles in straight ducts, specifically particles will focus towards stable equilibria located a small distance from the centre of the two longer walls.
For the smaller size particles one additionally finds stable equilibria near the centre of the side walls, however very few particles are likely to be found here in practice because the attractive region is much smaller~\citep{HoodThesis}.
Similar to the square duct case these generally only become evident for $R\gg10^{3}$ since even a small influence from the secondary flow is enough to modify/destroy these equilibria.
On the other hand, when $\kappa$ is large the effect of the secondary flow drag is dominant.
Unlike the square duct case, we find that the effect of the inertial lift force in the rectangular duct is enough that the centre of the vortices are stable equilibria.
However, it would be reasonable to expect this to break down for smaller particles, that is with $a<0.05$, as the relative influence of the secondary flow motion becomes more significant.
In the two limits discussed we note the stable equilibria pair is close to the centre horizontally.
For intermediate values of $\kappa$ on the other hand we observe the focusing position to be much closer to the inside wall.
The further left a particle can migrate varies with size and is between $0.3H$ and $0.4H$ for the radii considered herein.
Also noteworthy is that our results suggest that stable focusing positions can never occur in the right half of the cross-section.
In contrast, some experiments have shown focusing can occur in this region, although we believe this is likely to be an effect of higher flow rates and hence is not captured by the low flow rate approximation. 
In particular, at higher flow rates the location of the maximum of the axial flow and the centre of the secondary flow vortices are pushed towards the outer wall by inertia and it is reasonable to expect particle behaviour to be perturbed accordingly.

\begin{figure}
\centering
\begin{subfigure}[b]{\textwidth}
\centering
\includegraphics{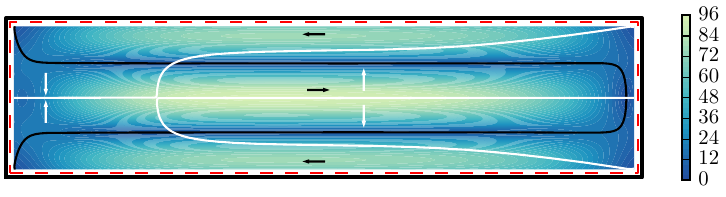}
\caption{$a=0.05$}
\end{subfigure}
\\
\begin{subfigure}[b]{\textwidth}
\centering
\includegraphics{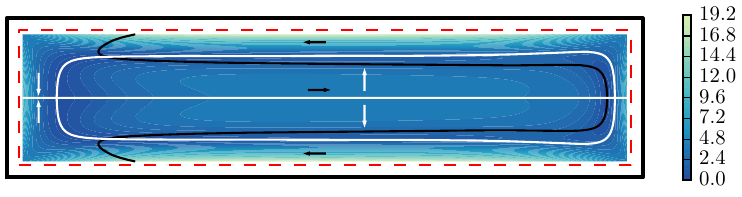}
\caption{$a=0.15$}
\end{subfigure}
\caption{
The force $\vec{F}_{p}^{\prime}$ on neutrally buoyant particles at different locations in a curved duct with rectangular cross-section ($W/4=H=2$) and bend radius $R=160$.
The colour background shows the magnitude of $\vec{F}_{p}^{\prime}$. 
Black and white contours are the zero level set curves of the horizontal and vertical components respectively whereas the arrows indicate the sign of each component in the area bounded by the respective contour.
The left wall is on the inside of the bend.
The dashed red line shows where the centre of the particle lies when its surface touches the wall.
}\label{fig:rect2_lift_plots_R160}
\end{figure}

\begin{figure}
\centering
\begin{subfigure}[b]{\textwidth}
\centering
\includegraphics{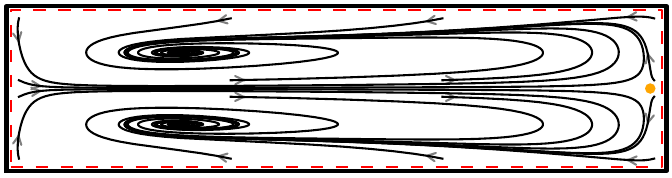}
\caption{$a=0.05$}
\end{subfigure}
\\
\begin{subfigure}[b]{\textwidth}
\centering
\includegraphics{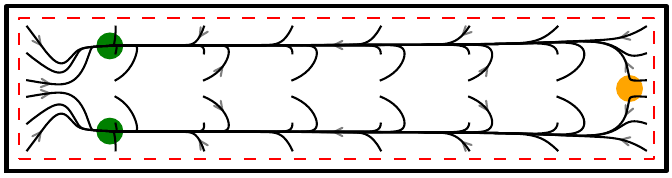}
\caption{$a=0.15$}
\end{subfigure}
\caption{
Approximate trajectories of neutrally buoyant particles in a curved duct with rectangular cross-section ($W/4=H=2$) and bend radius $R=160$.
Trajectories from several different starting positions have been super-imposed.
Green and orange  markers show the location of stable and (unstable) saddle equilibria respectively (with marker size indicative of particle size).
The left wall is on the inside of the bend.
The dashed red line shows where the centre of the particle lies when its surface touches the wall.
}\label{fig:rect2_traj_plots_R160}
\end{figure}

Force and trajectory plots for select particle sizes within the higher aspect ratio rectangular cross-section with $W=8$, $H=2$ and $R=160$  are shown in Figures~\ref{fig:rect2_lift_plots_R160} and \ref{fig:rect2_traj_plots_R160} respectively.
Generally speaking, the behaviour in the wider duct is qualitatively similar to that observed in the case with $W=4$.
This is evident in comparing Figures~\ref{fig:rect2_lift_plots_R160} and~\ref{fig:rect2_traj_plots_R160} with Figures~\ref{fig:rect_lift_plots_R160}(a,c) and~\ref{fig:rect_traj_plots_R160}(a,c) respectively.
The main advantage of the wider duct is a greater separation distance for particles focused towards the inside wall from those that focus towards the centre.
Variations in focusing position with respect to particle size and bend radius are critical in determining how effective a microfluidic duct is able to separate particles.
To better compare the differences for the two aspect ratios the focusing positions will be examined in detail.

%

\begin{figure}
\centering
\begin{subfigure}[b]{0.49\textwidth}
\centering
\includegraphics{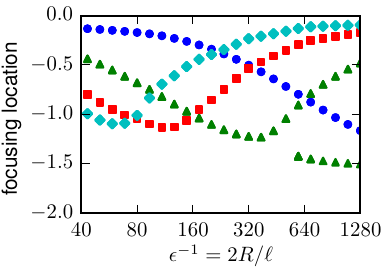} 
\caption{$W=4$ and $H=2$}
\end{subfigure}
\begin{subfigure}[b]{0.49\textwidth}
\centering
\includegraphics{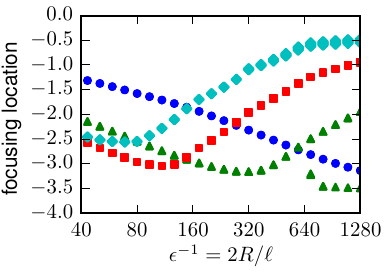} 
\caption{$W=8$ and $H=2$}
\end{subfigure}
\caption{
Horizontal location of the (stable) focusing positions of particles versus $\epsilon^{-1}=2R/\ell$ for the two rectangular ducts.
The particle sizes are $a=0.05$ (blue, circles), $a=0.10$ (green, triangles), $a=0.15$ (red, squares) and $a=0.20$ (cyan, diamonds).
Note the vertical axis includes only the left half of the duct ($0$ being the centre and $-2,-4$ being the inside wall in (a,b) respectively).
}\label{fig:R_focusing}
\end{figure}

To study how the focusing position of different size particles changes with respect to the bend radius, we interpolate $\vec{F}_{p}^{\prime}$ between $R\in\{640,320,160,80\}$ (and extrapolate for $R$ outside this range).
Since $\ell=2$ is fixed, changes in $R$ can be interpreted as changes to the dimensionless ratio $\epsilon^{-1}=2R/\ell$.
Despite the wide range of $R$, each of $\bar{u}_{\theta,0},\bar{u}_{r,0},\bar{u}_{z,0}$ (used to approximate $\hat{\bar{\vec{u}}}_{a},\hat{\bar{\vec{u}}}_{s}$ as in \eqref{eqn:bg_as_approx}) change only modestly. 
The most significant effect is a change in scale of the secondary flow drag relative to the inertial lift force, which is captured separately by the factor $\kappa=\ell^{4}/4a^{3}R$; changes in the remaining components of $\vec{F}_{p}^{\prime}$ are generally subtle.
For a much finer sampling of $R$ we then analyse the force on the particle $\vec{F}_{p}^{\prime}$ and identify/track the horizontal location of stable equilibria. 
The results of this analysis is shown in Figure~\ref{fig:R_focusing} for both rectangular ducts.
The vertical axis shows the horizontal location of the focusing position with the inside (left) wall at the bottom and the centre at the top.
The focusing position of particles undergoes a transition from being near the centre to being a finite distance away from the inside wall.
Importantly, this differs significantly with particle size, thereby suggesting a mechanism for size based separation of particles.
However, note that the relative ordering of particles changes many times over the range of $\epsilon^{-1}$ shown, suggesting that some care should be taken in choosing the bend radius depending on the size of particles to be separated.

For the duct with $W/H=2$ and with $\epsilon^{-1}\lessapprox80$ the ordering going from the inside wall to the centre is from largest to smallest particle.
As $\epsilon^{-1}$ increases there are many changes to the ordering as largest particles begin to migrate towards the centre whereas smaller ones migrate towards the wall.
Note that for $\epsilon^{-1}\gtrapprox500$ an additional stable equilibrium appears near the inside wall for the particle with radius $a=0.10$ and varies very little with respect to $\epsilon^{-1}$.
This is one of the extra equilibria that is expected to occur as the behaviour approaches that of a straight duct.
For $\epsilon^{-1}\gtrapprox640$ the ordering of the dominant equilibria is from smallest to largest particle.
Similar trends are observed for the duct with $W/H=4$.
In particular, the ordering of the focusing location for $\epsilon^{-1}\lessapprox80$ is the same with the exception of the two larger particles being switched.
There are then several changes to the ordering for increasing $\epsilon^{-1}$ until for $\epsilon^{-1}\gtrapprox640$ the ordering is from smallest to largest particle (noting we again observe the emergence of an additional stable equilibrium for $a=0.15$ near the inside wall).
One notable difference for the two different aspect ratios is that particles generally don't appear to get as close to the centre over the range of $\epsilon^{-1}$ shown.

\begin{figure}
\centering
\begin{subfigure}[b]{0.49\textwidth}
\centering
\includegraphics{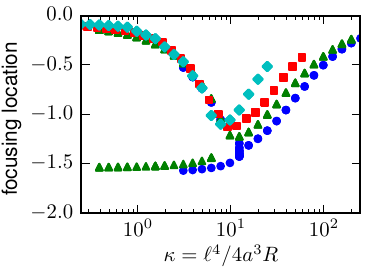}
\caption{$W=4$ and $H=2$}
\end{subfigure}
\begin{subfigure}[b]{0.49\textwidth}
\centering
\includegraphics{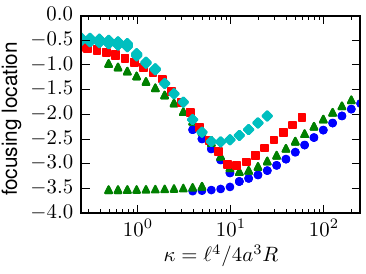}
\caption{$W=8$ and $H=2$}
\end{subfigure}
\caption{
Horizontal location of the (stable) focusing positions of particles versus $\kappa=\ell^{4}/4a^{3}R$ for the two rectangular ducts.
The particle sizes are $a=0.05$ (blue, circles), $a=0.10$ (green, triangles), $a=0.15$ (red, squares) and $a=0.20$ (cyan, diamonds).
Note the vertical axis includes only the left half of the duct.
}\label{fig:kappa_focusing}
\end{figure}

Figure~\ref{fig:kappa_focusing} shows the same data plotted against the ratio $\kappa=\ell^{4}/4a^{3}R$ for both rectangular ducts.
Recall that small $\kappa$ indicates that inertial lift forces are dominant and large $\kappa$ indicates that secondary flow drag is dominant.
The focusing horizontal focusing location approximately collapses onto a single curve, particularly for $\kappa\ll10$, in each case.
Note that for the two smaller particle sizes ($a\in\{0.05,0.10\}$) an additional stable equilibrium is observed when $\kappa$ is small (these are the additional `tails' to the main curve).
These are again expected, since $R$ is comparatively large such that the focusing behaviour is approaching that of a straight duct, and are not part of the scaling theory.
Going from small to large $\kappa$ we can see that particles initially focus close to the centre of the duct, then move towards the left wall before reaching a minimum and then gradually moving back towards the centre again.
Observe that the minimum achieved is increasing with respect to particle size, that is larger particles do not get as close to the inside wall as smaller ones.
This effect is amplified in the case of the largest particle in the wider of the two ducts.

Experiments using a spiral duct with two different rectangular cross-sections were performed by \citet{WuEtal2012}.
Both ducts had a width of $W=500\mu$m but with different heights $H=90\mu$m and $H=120\mu$m.
Beads with diameter $2a=10\mu$m and $2a=6\mu$m were suspended in flow through the ducts at flow rates of $1$mL$/$min and $2$mL$/$min for the smaller and larger cross-sections respectively.
In both cases the larger particles focused towards the inside wall whereas the smaller ones were less focused in a region centred slightly towards the inside wall from the centre. 
The corresponding $\kappa$ (near the outlet where $R$ is smallest) for the larger particle is approximately $29$ and $92$ for the smaller and larger cross-section respectively, while for the smaller particle it is approximately $135$ and $427$. 
Looking at the results in Figure~\ref{fig:kappa_focusing}(b), noting that the particle sizes $a=0.05,0.10$ are closest to those of the experiment (given $H=2$ in our computations), we see that the larger particle is indeed expected to focus nearer the inside wall in each case.
Thus our model shows qualitative agreement with these experiments.

We summarise our results here for the rectangular ducts.
With the exception of an additional focusing position for small $a$ and large $R$, there is generally a unique horizontal focusing location of a particle suspended in flow through a curved rectangular duct.
This is significantly different from the square duct case and explains why rectangular ducts are particularly useful for microfluidic applications requiring a focused stream of particles.
Furthermore, the horizontal focusing position is sensitive to both the particle size and bend radius, providing a practical mechanism for sized based particle separation.
However, our results demonstrate that the ordering of focused particles changes several times with bend radius even in the simplified scenario of low flow rates. 
This suggests that much care should be taken with device design, specifically the choice of bend radius, depending upon the size of particles to be separated. 
We have shown that the horizontal focusing position approximately collapses onto a single curve when plotted against $\kappa=\ell^{4}/4a^{3}R$.
Additionally, although there are some small differences for the two different aspect ratio ducts considered, the general focusing behaviour is qualitatively similar with respect to $\kappa$.
This provides a tangible variable that can be used to guide the design of devices with rectangular cross-section operating at suitably low flow rates.

\subsection{Ducts having a trapezoidal cross-section}

\begin{figure}
\centering
\begin{subfigure}[b]{\textwidth}
\centering
\includegraphics{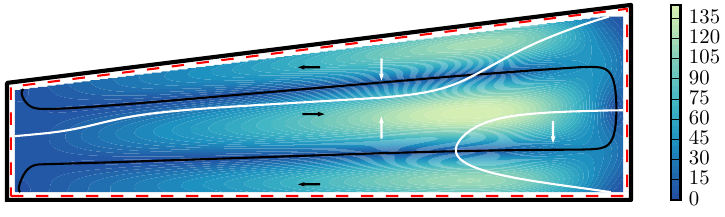}
\caption{$a=0.05$}
\end{subfigure}
\\
\begin{subfigure}[b]{\textwidth}
\centering
\includegraphics{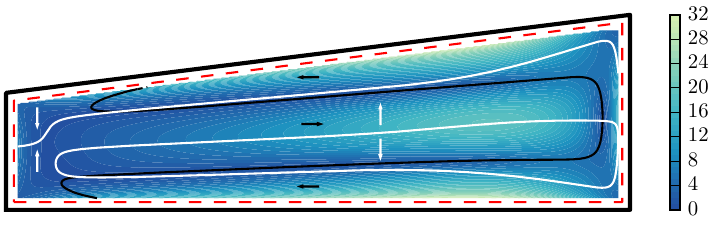}
\caption{$a=0.10$} 
\end{subfigure}
\\
\begin{subfigure}[b]{\textwidth}
\centering
\includegraphics{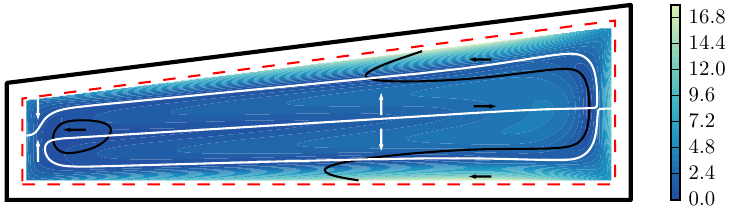}
\caption{$a=0.20$}
\end{subfigure}
\caption{
The force $\vec{F}_{p}^{\prime}$ on neutrally buoyant particles in a curved duct with trapezoidal cross-section ($W=8$, $H_{\text{left}}=3/2$ and $H_{\text{right}}=5/2$) and bend radius $R=160$.
The colour background shows the magnitude of $\vec{F}_{p}^{\prime}$. 
Black and white contours are the zero level set curves of the horizontal and vertical components respectively whereas the arrows indicate the sign of each component in the area bounded by the respective contour.
The left wall is on the inside of the bend.
The dashed red line shows where the centre of the particle lies when its surface touches the wall.
}\label{fig:trap_lift_plots_R160}
\end{figure}

\begin{figure}
\centering
\begin{subfigure}[b]{0.9\textwidth}
\centering
\includegraphics{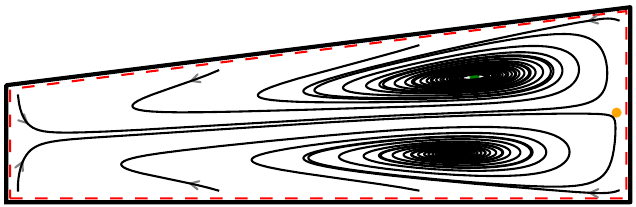}
\caption{$a=0.05$}
\end{subfigure}
\\
\begin{subfigure}[b]{0.9\textwidth}
\centering
\includegraphics{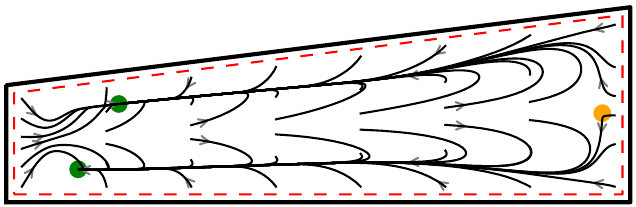}
\caption{$a=0.10$} 
\end{subfigure}
\\
\begin{subfigure}[b]{0.9\textwidth}
\centering
\includegraphics{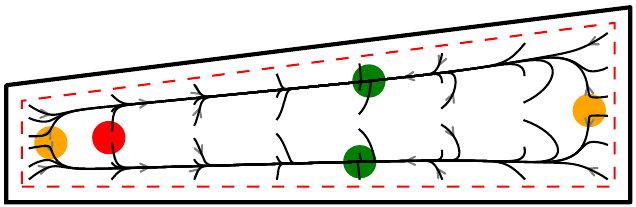}
\caption{$a=0.20$}
\end{subfigure}
\caption{
Approximate trajectories of neutrally buoyant particles in a curved duct with trapezoidal cross-section ($W=8$, $H_{\text{left}}=3/2$ and $H_{\text{right}}=5/2$) and bend radius $R=160$.
Trajectories from several starting positions are super-imposed.
Green, orange and red markers show the location of stable, (unstable) saddle and unstable equilibria respectively (with marker size indicative of particle size).
The left wall is on the inside of the bend.
The dashed red line shows where the centre of the particle lies when its surface touches the wall.
}\label{fig:trap_traj_plots_R160}
\end{figure}

We now consider the behaviour of particles in a curved duct with trapezoidal cross-section inspired by the experiments of \citet{GuanEtal2013} and \citet{Warkianietal2014} in which the bottom wall remains perpendicular to the flow direction and parallel to the bend plane, whilst the top wall is sloped (with straight/vertical side walls).
We consider a cross-section width $W=8$ and heights $H_{\text{left}}=1.5$ and $H_{\text{right}}=2.5$ at the left (inside) and right (outside) walls respectively (see figures below). 
This new shape has a significant effect on the background flow compared to the rectangular ducts, in particular the location of the maximum of the axial component and the centre of the vortices of the secondary component are each shifted towards the outside wall.
Furthermore, the asymmetry of the cross-section means that the axial and secondary components no longer have even and odd symmetry respectively with respect to $z$.
These changes have a significant effect on the focusing behaviour of particles within the duct.

The perturbing force $\vec{F}_{p}^{\prime}$ was computed for each particle size $a\in\{0.20,0.15,0.10,0.05\}$ and bend radius $R\in\{640,320,160,80\}$. 
Figures \ref{fig:trap_lift_plots_R160} and \ref{fig:trap_traj_plots_R160} show force and trajectory plots respectively in the specific case of $R=160$ for the particle sizes $a\in\{0.05,0.10,0.20\}$.
Results for the smallest particle ($a=0.05$ and $\ell^{4}/4a^{3}R=200$) are shown in Figures~\ref{fig:trap_lift_plots_R160}(a) and \ref{fig:trap_traj_plots_R160}(a).
It is immediately evident from the zero level set curves that the asymmetric shape of the cross-section has a significant impact on the location of equilibria in comparison to the rectangular duct case.
Three equilibria can be identified in the force plot, one very close to the centre of the outer wall which is clearly a saddle equilibrium.
The trajectory plot shows that, despite the secondary flow drag being the dominant effect, the particles gradually spiral in towards one of the two remaining equilibria which are stable (the green marker is somewhat hidden by the trajectories).
In contrast to the results for a rectangular duct, these stable equilibria are in the outside half of the cross-section and are also slightly staggered with respect to their horizontal position.

The results for the second smallest particle ($a=0.10$ and $\ell^{4}/4a^{3}R=25$) are shown in Figures~\ref{fig:trap_lift_plots_R160}(b) and \ref{fig:trap_traj_plots_R160}(b).
Three equilibria can again be identified, a saddle remains near the outside wall whereas the stable pair has shifted to the opposite side of the cross-section.
The magnitude of the secondary flow drag and inertial lift force is similar and their interaction leads to this significant shift in the focusing location.
The particle dynamics shows an initial converge onto a `slow manifold' along which a slower migration towards one of the two stable equilibria takes place.
Note that the horizontal location of the stable equilibria is staggered even more so than in the case of the smaller particle.
The force and trajectories of the largest particle ($a=0.20$ and $\ell^{4}/4a^{3}R=3.125$) are shown in Figures~\ref{fig:trap_lift_plots_R160}(c) and \ref{fig:trap_traj_plots_R160}(c).
Five equilibria can be identified in this case. 
Three of these lie on the white contour running along the centre (vertically) and are all unstable (the outer two are saddles).
The remaining two are stable and are now located slightly right of centre. 
The inertial lift is dominant in this case and the trajectory plot demonstrates that particles converge more directly onto the `slow manifold' before migrating towards a stable equilibrium.
The results for $a=0.15$ have been omitted as they are similar to the case $a=0.20$ but with the stable pair slightly left of centre.

\begin{figure}
\centering
\begin{subfigure}[b]{0.48\textwidth}
\centering
\includegraphics{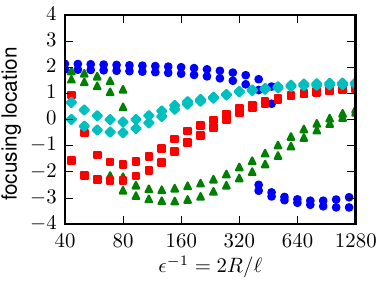}
\caption{}
\end{subfigure}
\quad
\begin{subfigure}[b]{0.48\textwidth}
\centering
\includegraphics{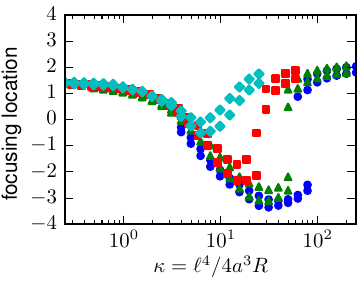}
\caption{}
\end{subfigure}
\caption{
Horizontal location of (stable) focusing positions of particles with respect to (a) $\epsilon^{-1}=2R/\ell$ and (b) $\kappa\ell^{4}/4a^{3}R$ for a trapezoidal duct with $W=8$, $H_{\text{left}}=3/2$ and $H_{\text{right}}=5/2$.
The particle sizes are $a=0.05$ (blue, circles), $a=0.10$ (green, triangles), $a=0.15$ (red, squares) and $a=0.20$ (cyan, diamonds).
}\label{fig:trap_focusing}
\end{figure}

We again consider how the focusing location changes with respect to $R$ (or equivalently $\epsilon^{-1}=2R/\ell$ given fixed $\ell$).
Using the same approach as in the case of the rectangular cross-sections, the $\vec{F}_{p}^{\prime}$ are interpolated between the given $R$ samples and the horizontal location of the stable equilibria are determined over a finer sampling of $R$.
The horizontal focusing position versus $\epsilon^{-1}$ is shown in Figure~\ref{fig:trap_focusing}(a).
The staggering of the equilibria pair due to the asymmetry of the cross-section is shown as two different, albeit close, focusing positions for each $a$ and $R$.
In Figure~\ref{fig:trap_focusing} (a) there are three main regions that can be identified over the $\epsilon^{-1}$ considered. 
For $\epsilon^{-1}\lessapprox80$ the smaller two particle sizes are focused near $r=2$, the largest near $r=0$, and the second largest shows some significant staggering but is generally on the inside half of the cross-section. 
For $80\lessapprox \epsilon^{-1}\lessapprox400$ the second smallest particle effectively jumps and now focuses nearest to the inside wall while the relative order of the remaining three is the same.
For $R\gtrapprox 400$ the smallest particle has now jumped and focuses nearest to the inside wall.
These apparent discontinuities in the focusing location with respect to bend radius are a significant difference compared to the rectangular duct case where the curve is smooth.

\begin{figure}
\centering
\begin{subfigure}[b]{\textwidth}
\centering
\includegraphics{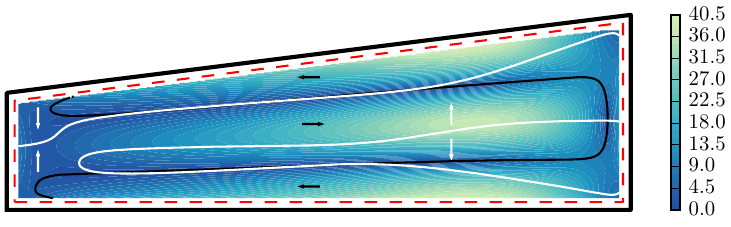}
\caption{Force plot}
\end{subfigure}
\\
\begin{subfigure}[b]{\textwidth}
\centering
\includegraphics{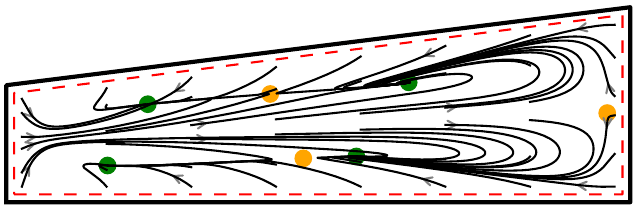}
\caption{Trajectory plot}
\end{subfigure}
\caption{
The force $\vec{F}_{p}^{\prime}$ and trajectories of a neutrally buoyant particle having radius $a=0.10$ in a curved duct having trapezoidal cross-section ($W=8$, $H_{\text{left}}=3/2$ and $H_{\text{right}}=5/2$) and bend radius $R=80$.
Interpretation of the two plots is the same as in Figures~\ref{fig:trap_lift_plots_R160} and~\ref{fig:trap_traj_plots_R160} respectively.
Many stable equilibria can be observed in this case.
}\label{fig:trap_plots_a10_R80}
\end{figure}

Looking carefully around the discontinuities in Figure~\ref{fig:trap_focusing}(a) there is a small range for which more than two stable focusing positions exist. 
Specifically, this is evident for the smallest particle ($a=0.05$, blue circles) at $\epsilon^{-1}\approx400$ and the second smallest particle ($a=0.10$, green triangles) at $\epsilon^{-1}\approx 80$.
The force and trajectory plots for the latter case ($a=0.10$ and $\epsilon^{-1}=80$) is shown in Figure~\ref{fig:trap_plots_a10_R80}.
Comparing Figure~\ref{fig:trap_plots_a10_R80}(a) with Figure~\ref{fig:trap_lift_plots_R160}(b) the white contours along the mid-section have shifted relative to the black contours such that they each now cross three times resulting in two stable equilibria and an unstable saddle in between.
With further decrease in $\epsilon^{-1}$ (or equivalently $R$) the relative location of the black and white contours in the mid-section shift further such that only the right most intersections remain thus leaving only the two stable equilibria on the right.
The trajectories in Figure~\ref{fig:trap_plots_a10_R80}(b) suggest that the majority of starting positions migrate towards the two stable equilibria on the right.
However, in the context of a spiral device in which $R$ is continuously changing, this may not be the case and further investigation is required.

In Figure~\ref{fig:trap_focusing} (b) the focusing data is plotted against the ratio $\kappa=\ell^{4}/4a^{3}R$.
In a broad sense the general trend is similar to the rectangular duct case, specifically the focusing positions move towards the inside wall and back again with increasing $\kappa$.
Additionally, for $\kappa<10$ and $\kappa>100$ the behaviour approximately collapses onto a single curve. 
However, for intermediate $\kappa$ some significant differences in focusing position occur depending on the particle size.
The three smaller particle sizes achieve a stable focusing position within two units of the left wall for $\kappa$ in this intermediate range. 
Each also exhibits a rapid change in focusing position from the inside half of the duct to the outside half when $\kappa$ is a little larger than where the minimum is achieved.
Furthermore, this rapid change occurs at larger $\kappa$ as the particle size decreases.
On the other hand, the largest particle ($a=0.20$) does not get close to the inside wall for any $\kappa$ (and is only briefly left of centre).
Furthermore, its focusing behaviour appears to be reasonably smooth with respect to $\kappa$ compared to that observed for the smaller particles.

It is interesting to consider how these different features of focusing behaviour can affect the efficiency of such devices for the sized based separation of particles.
The staggering of the horizontal location of the stable equilibria in the top half and bottom half of the duct is somewhat undesirable.
To some degree this is offset by the larger separation distances provided by the trapezoidal duct, resulting from stable equilibria being able to exist in the outer half of the cross-section.
An obvious modification of the duct would be to make the bottom wall slanted (in the opposite direction) to achieve a symmetric trapezoidal shape.
This should eliminate the horizontal staggering of equilibria pairs while maintaining a large separation distance, although it may be difficult to reliably produce such cross-sections.
We also observe that the focusing location of smaller particles can change rapidly with respect to small changes in $R$. 
While this results in a narrow band of design parameters in which the particle may be found at multiple locations, it could also be exploited in carefully designed microfluidic devices to achieve reasonably high separation efficiency of particles whose size may differ by a comparatively small amount.

Experiments in two different spiral ducts with trapezoidal cross-section using beads with diameter $2a=10\mu$m and $2a=6\mu$m were performed by \citet{WuEtal2012}.
Both ducts had a width of $W=500\mu$m but with the different heights of $H_{\text{left}}=70\mu$m to $H_{\text{right}}=100\mu$m from the inner to outer wall in the smaller cross-section and $H_{\text{left}}=90\mu$m to $H_{\text{right}}=120\mu$m in the larger cross-section. 
In both cases the larger particles focused towards the inside wall while the smaller ones were less focused in a region slightly towards the outside wall from the centre. 
The corresponding $\kappa$ (near the outlet where $R$ is smallest) for the larger particle is approximately $23$ and $54$ for the smaller and larger cross-section respectively, while for the smaller particle it is approximately $107$ and $250$. 
Looking at the results in Figure~\ref{fig:trap_focusing}(b), noting that the particle sizes $a=0.05,0.10$ are closest to those of the experiment, we see that the larger particle is indeed expected to focus near the inside wall for these $\kappa$ whereas the smaller particle will focus towards the outer wall.
Interestingly the staggering of the equilibria pair is not evident in the experimental results although this is potentially because the staggering is less than the width of the fluorescent streaks in the experimental data.
%

\section{Conclusions}


This paper develops a general model for the forces that govern the motion of a spherical particle suspended in flow through a curved duct.
A key component in extending the approach of \citet{HoodLeeRoper2015} from straight ducts to curved ducts is the use of a rotating coordinate system as a frame of reference in which the flow is approximately steady.
Additionally, an expansion of the background flow into axial and secondary components identifies how different components of the force on a particle are affected by the background flow.
We performed further analysis on the special case of low flow rate and neutrally buoyant particles.
We found that the the secondary flow drag scales with $\kappa=\ell^{4}/4a^{3}R$ relative to the inertial lift force.
We computed the position and stability of equilibria for several different cross-sectional geometries.
Further, a simple first order model of particle trajectories allows us to plot approximate trajectories and identify a slow manifold in many cases.

An analysis of the location of stable equilibria in rectangular and trapezoidal cross-sections demonstrates that $\kappa$ plays an important role in the general focusing behaviour of particles. 
We observed that a stable equilibria pair exists over a large range of bend radii and particle diameter.  
While several changes in the lateral ordering of focused particles are observed we find that the focusing behaviour approximately collapse onto a single curve when plotted against $\kappa$, particularly for rectangular cross-sections.
The results suggest that this is the mechanism for the size based particle/cell separation observed in the experimental literature, providing a dimensionless parameter that may aid in the design and operation of microfluidic ducts under appropriate conditions.
%

While our study assumes the bend radius to be constant the results can be applied to provide some insight into the focusing behaviour within spiral microfluidic ducts.
In particular, the final location of a particle in a spiral device can be estimated by looking at the focusing behaviour in a curved duct with (constant) bend radius matching that near the outlet of the spiral.
Furthermore, our results that illustrate the change in particle focusing position with respect to $\epsilon^{-1}=2R/\ell$ illustrate where particles will focus towards as the bend radius changes throughout a spiral device.
Lastly we note that the results could be reasonably applied to dilute suspensions of particles/cells provided the concentration is small enough that interactions between particles/cells can be neglected and the particles/cells are sufficiently rigid and sphere like in shape.

There are several ways in which this work could be extended.
This study considered a single trapezoidal shaped duct and an interesting extension is to examine how the focusing dynamics evolve as the slope of the top wall is slowly increased starting from a rectangular cross-section.
Further, one could investigate what happens when the trapezoidal cross-section is tallest at the inside wall as in the experiment of \citet{SofelaEtal2018}.
Exploration of other cross-sectional geometries may also be interesting, for example a symmetric trapezoidal shape which we expect to eliminate the staggering of equilibria observed with the asymmetric trapezoidal shape.
One could even go further and consider cross-sections in which the top and bottom walls have a more general polynomial shape.

It would be interesting to explore how a change in density of the particle influences the existence and stability of equilibria. 
In particular, if the density of the particle differs enough from that of the fluid that gravity becomes relevant then this will break the symmetry of the force on the particles in rectangular cross-sections.
Additionally, there will be a noticeable change in the centrifugal force for a non-neutrally buoyant particle which will influence the horizontal location of equilibria.
The focusing behaviour at higher flow rates is another important aspect to consider in order to achieve a reasonable throughput while maintaining a high separation efficiency.
Finally, the simple first order trajectory model might be extended to a more complete second order model which takes into account axial acceleration/deceleration of the particle as it migrates within the cross-section.
In particular it is reasonable to expect changes to the axial velocity to have an appreciable effect on the inertial lift force, particularly when $\kappa$ is large.

\section*{Acknowledgements}
This research was supported by the Australian Research Council via a Discovery Project (DP160102021) and a Future Fellowship (FT160100108) to YMS, and by the Simons Foundation via a Math + X grant (510776) to ALB.
Funding from the University of Adelaide via an establishment grant to YMS is also gratefully acknowledged.
The results were computed using supercomputing resources provided by the Phoenix HPC service at the University of Adelaide.


\appendix

\section{Nomenclature}

\begin{table}
\caption{Summary of nomenclature used throughout the paper.}\label{tab:notation}
\begin{tabular}{ll}
\hline
symbol & description \\
\hline
$a$ & radius of the (spherical) particle \\
$\mathcal{D},\partial\mathcal{D}$ & duct interior and its boundary (in lab frame) \\
$\vec{F}$ & force on a particle (in lab frame) \\
$\vec{F}_{p}^{\prime}$ & cross-sectional forces on particle from our model \\
$\mathcal{F},\partial\mathcal{F}$ & fluid domain and its boundary when a particle is present (in lab frame) \\
$G$ & pressure gradient along duct centre-line which drives the fluid flow \\
$H$ & height of the duct cross-section \\
$\ell$ & characteristic dimension of duct cross-section (taken to be $\min\{H,W\}$) \\
$I_{p}$ & moment of inertia of the (spherical) particle \\
$\mathbb{I}$ & identity tensor \\
$K$ & perturbation parameter of the background fluid flow (taken to be $\epsilon\Re^{2}/4$) \\
$m_{p}$ & mass of the (spherical) particle \\
$\vec{n}$ & normal vector on a prescribed surface pointing outwards relative to the fluid \\
$p,\bar{p}$ & fluid pressure with and without particle present respectively (in lab frame) \\
$q$ & disturbance fluid pressure \\
$r$ & radial coordinate with respect to the $x$--$y$ plane (in the lab frame) \\
$r_{p}$ & radial coordinate of particle position (frame independent) \\
$R$ & bend radius of the duct (measured to centre of cross-section) \\
$\Re$ & Reynolds number (taken to be $(\rho/\mu)U_{m}\ell$) \\
$\Rep$ & particle Reynolds number (taken to be $(a/\ell)^{2}\Re$) \\
$t$ & time variable \\
$\vec{T}$ & torque on a particle (in lab frame) \\
$\vec{u},\bar{\vec{u}}$ & fluid velocity with and without particle present respectively (in lab frame) \\
$\bar{\vec{u}}_{a},\bar{\vec{u}}_{s}$ & axial and secondary components of the background fluid velocity \\
$\vec{u}_{p}$ & particle velocity (in lab frame) \\
$U_{m}$ & characteristic velocity of background flow (taken as the max of $\bar{\vec{u}}_{a}$) \\
$U_{p}$ & axial component of the particle velocity \\
$\vec{v}$ & disturbance fluid velocity \\
$W$ & width of duct cross-section \\
$\vec{x}$ & spatial coordinates (in lab frame) \\
$\vec{x}_{p}$ & particle position (in lab frame)  \\
$x_{p},y_{p},z_{p}$ & Cartesian coordinates of particle position (in lab frame) \\ 
%
$\epsilon$ & half the characteristic duct length divided by the bend radius, that is $\ell/2R$ \\
$\theta$ & angular coordinate with respect to the $x$--$y$ plane (in lab frame) \\ 
$\theta_{p}$ & angular coordinate of particle position (in lab frame) \\
$\Theta$ & angular velocity of the rotating reference frame ($z$ component) \\
$\kappa$ & dimensionless scale of the secondary flow drag (relative to the inertial lift force) \\
$\mu$ & fluid viscosity \\
$\rho$ & fluid density \\
$\rho_{p}$ & particle density (taken to be uniform) \\
$\vec{\Omega}_{p}$ & particle spin (in lab frame) 
%
\end{tabular}
\vspace{2.0cm} 
\end{table}

Table~\ref{tab:notation} summarises the nomenclature used throughout the paper.
We note that a large number of variables listed are also introduced in the context of a rotating reference frame which is always denoted with a prime (e.g. as in $\ast'$).
For brevity these variants are excluded from the table.
Dimensionless forms of variables are denoted throughout the paper with a hat/caret (i.e. as in $\hat{\ast}$) and are omitted from the table for brevity.
Perturbation expansions are also introduced throughout the paper via a subscript, for example $\hat{\vec{v}}'=\vec{v}_{0}+\Rep\vec{v}_{1}+O(\Rep^{2})$, and are also omitted from the table.
We additionally note that the caret and prime are dropped from perturbation variables for ease of readability since these are always taken to refer to dimensionless quantities in the rotating reference frame.
Lastly, some terms are decomposed into separate parts dependent on the axial and secondary components of the background flow typically denoted as $\ast_{a},\ast_{s}$.
With the exception of $\bar{\vec{u}}_{a},\bar{\vec{u}}_{s}$ these too are omitted from the table.

\section{Estimation of the background flow}\label{app:bgf}

We provide a brief account of the derivation of the equations governing the background flow from \citet{Harding2018} which we utilise in Sections~\ref{sec:lfr_limit} and \ref{sec:results}.
In the notation of Section~\ref{sec:gov_eqns} the Navier--Stokes equations for steady flow through a curved duct in cylindrical coordinates are 
\begin{subequations}\label{eqn:bg_NS}
\begin{align}
0 &= \frac{\partial \bar{u}_{r}}{\partial r}+\frac{\partial \bar{u}_{z}}{\partial z}+\frac{\bar{u}_{r}}{R+r}  , \\
\rho\left(\bar{u}_{r}\frac{\partial \bar{u}_{\theta}}{\partial r}+\bar{u}_{z}\frac{\partial \bar{u}_{\theta}}{\partial z}+\frac{\bar{u}_{\theta}\bar{u}_{r}}{R+r}\right) &= \frac{GR}{R+r}+\mu\left(\frac{\partial^{2} \bar{u}_{\theta}}{\partial r^{2}}+\frac{\partial^{2} \bar{u}_{\theta}}{\partial z^{2}}+\frac{1}{R+r}\frac{\partial \bar{u}_{\theta}}{\partial r}-\frac{\bar{u}_{\theta}}{(R+r)^{2}}\right)  , \\
\rho\left(\bar{u}_{r}\frac{\partial \bar{u}_{r}}{\partial r}+\bar{u}_{z}\frac{\partial \bar{u}_{r}}{\partial z}-\frac{\bar{u}_{\theta}^{2}}{R+r}\right) &= -\frac{\partial \tilde{p}}{\partial r}+\mu\left(\frac{\partial^{2} \bar{u}_{r}}{\partial r^{2}}+\frac{\partial^{2} \bar{u}_{r}}{\partial z^{2}}+\frac{1}{r}\frac{\partial \bar{u}_{r}}{\partial r}-\frac{\bar{u}_{r}}{r^{2}}\right)  , \label{eqn:NS_c} \\
\rho\left(\bar{u}_{r}\frac{\partial \bar{u}_{z}}{\partial r}+\bar{u}_{z}\frac{\partial \bar{u}_{z}}{\partial z}\right) &= -\frac{\partial \tilde{p}}{\partial z}+\mu\left(\frac{\partial^{2} \bar{u}_{z}}{\partial r^{2}}+\frac{\partial^{2} \bar{u}_{z}}{\partial z^{2}}+\frac{1}{r}\frac{\partial \bar{u}_{z}}{\partial r}\right)  .
\end{align}
\end{subequations}
Solving these equations is most easily done via a non-dimensionalisation that reflects the scaling of the background flow rather than the flow near a particle.
In particular, the spatial coordinates are non-dimensionalised as $(r,z)=(\ell/2)(\grave{r},\grave{z})$ and the fluid velocity components are non-dimensionalised as $\bar{u}_{\theta}=U_{m}\grave{\bar{u}}_{\theta}$ and $(\bar{u}_{r},\bar{u}_{z})=(\epsilon\Re/2) U_{m}(\grave{\bar{u}}_{r},\grave{\bar{u}}_{z})$. 
Observe that $R+r=R(1+\epsilon\grave{r})$ where $\epsilon=\ell/2R$ and for convenience we define $\grave{R}:=(1+\epsilon\grave{r})$ such that $R+r=R\grave{R}$.
We additionally introduce the (dimensionless) stream function $\Phi$ such that $\partial\Phi/\partial \grave{z}=-\grave{R}\grave{\bar{u}}_{r}$ and $\partial\Phi/\partial \grave{r}=-\grave{R}\grave{\bar{u}}_{z}$.
The equations \eqref{eqn:bg_NS} can then be reduced to
\begin{align*}
&K\left(-\frac{\partial \Phi}{\partial \grave{z}}\frac{\partial \grave{\bar{u}}_{\theta}}{\partial \grave{r}}+\frac{\partial \Phi}{\partial \grave{r}}\frac{\partial \grave{\bar{u}}_{\theta}}{\partial \grave{z}}-\epsilon\frac{\grave{\bar{u}}_{\theta}}{\grave{R}}\frac{\partial \Phi}{\partial \grave{z}}\right) 
= G+\grave{R}\Delta \grave{\bar{u}}_{\theta}+\epsilon\frac{\partial \grave{\bar{u}}_{\theta}}{\partial \grave{r}}-\epsilon^{2}\frac{\grave{\bar{u}}_{\theta}}{\grave{R}}  , \\
&K\left(\epsilon\frac{2}{\grave{R}^{3}}\frac{\partial^{2}\Phi}{\partial \grave{z}^{2}}\frac{\partial\Phi}{\partial \grave{z}}-\frac{1}{\grave{R}^{2}}\frac{\partial\Phi}{\partial \grave{z}}\frac{\partial\Delta\Phi}{\partial \grave{r}}+\frac{1}{\grave{R}^{2}}\frac{\partial \Phi}{\partial \grave{r}}\frac{\partial \Delta\Phi}{\partial \grave{z}}-\epsilon^{2}\frac{3}{\grave{R}^{4}}\frac{\partial\Phi}{\partial \grave{z}}\frac{\partial\Phi}{\partial \grave{r}}  +\epsilon\frac{3}{\grave{R}^{3}}\frac{\partial\Phi}{\partial \grave{z}}\frac{\partial^{2}\Phi}{\partial \grave{r}^{2}} \right. \notag\\
&\qquad\left.-\epsilon\frac{1}{\grave{R}^{3}}\frac{\partial\Phi}{\partial \grave{r}}\frac{\partial^{2}\Phi}{\partial \grave{r}\partial \grave{z}}\right) 
+\frac{2\grave{\bar{u}}_{\theta}}{\grave{R}}\frac{\partial \grave{\bar{u}}_{\theta}}{\partial \grave{z}}= \frac{1}{\grave{R}}\Delta^{2}\Phi -\epsilon\frac{2}{\grave{R}^{2}}\frac{\partial\Delta\Phi}{\partial \grave{r}}+\epsilon^{2}\frac{3}{\grave{R}^{3}}\frac{\partial^{2}\Phi}{\partial \grave{r}^{2}}-\epsilon^{3}\frac{3}{\grave{R}^{4}}\frac{\partial \Phi}{\partial \grave{r}} ,
\end{align*}
where $K=\epsilon\Re^{2}/4$ and $\Delta:=\partial^{2}/\partial\grave{r}^{2}+\partial^{2}/\partial\grave{z}^{2}$.
For small $K$ we can introduce a perturbation expansion of $\grave{\bar{u}}_{\theta}$ and $\Phi$ with respect to $K$, specifically
\begin{equation*}
\grave{\bar{u}}_{\theta}=\sum_{i=0}^{\infty}K^{i}\grave{\bar{u}}_{\theta,i} , \qquad \Phi=\sum_{i=0}^{\infty}K^{i}\Phi_{i} .
\end{equation*}
For $K\ll1$ it is sufficient to estimate the cross-sectional forces on a particle using the approximation $\grave{\bar{u}}_{\theta}\approx\grave{\bar{u}}_{\theta,0}$ and $\Phi\approx\Phi_{0}$ where
\begin{align*}
-G&=\grave{R}\Delta \grave{\bar{u}}_{\theta,0}+\epsilon\frac{\partial \grave{\bar{u}}_{\theta,0}}{\partial \grave{r}}-\epsilon^{2}\frac{\grave{\bar{u}}_{\theta,0}}{\grave{R}}  , \\
\frac{2\grave{\bar{u}}_{\theta,0}}{\grave{R}}\frac{\partial \grave{\bar{u}}_{\theta,0}}{\partial \grave{z}}&= \frac{1}{\grave{R}}\Delta^{2}\Phi_{0} -\epsilon\frac{2}{\grave{R}^{2}}\frac{\partial\Delta\Phi_{0}}{\partial \grave{r}}+\epsilon^{2}\frac{3}{\grave{R}^{3}}\frac{\partial^{2}\Phi_{0}}{\partial \grave{r}^{2}}-\epsilon^{3}\frac{3}{\grave{R}^{4}}\frac{\partial \Phi_{0}}{\partial \grave{r}} .
\end{align*}
A Rayleigh--Ritz method for approximating these components, which is used in the numerics in Section~\ref{sec:results}, is described in~\citet{Harding2018}.


\end{document}